\documentclass[onecolumn]{aa} % for a paper on 1 column 

\usepackage[T1]{fontenc}
\usepackage[latin1]{inputenc}
\usepackage{wasysym}
\usepackage{graphicx}
\usepackage{natbib}
\usepackage{txfonts}

\newcommand{\eq}[1]{\begin{equation} #1 \end{equation}}
\newcommand{\eqnn}[1]{\[ #1 \]}
\newcommand{\eqa}[1]{\begin{eqnarray} #1 \end{eqnarray}}

\newcommand{\br}[1]{\left( #1 \right)}
\newcommand{\bc}[1]{\left\{ #1 \right\}}

\newcommand{\ba}[1]{\left\langle #1 \right\rangle}
\newcommand{\bs}[1]{\left| #1 \right|}

\newcommand{\nn}{\nonumber}

\newcommand{\dd}{{\rm d}}
\newcommand{\vek}[1]{\mbox{\boldmath $#1$}}
\newcommand{\svek}[1]{\mbox{\boldmath \scriptsize $#1$}}  %small vectors
\newcommand{\ic}{{\rm i}}

\begin{document}

%Title
\title{Analysis of two-point statistics of cosmic shear}
\subtitle{III. Covariances of shear measures made easy}

\author{B. Joachimi, P. Schneider, \and T. Eifler %\inst{1}
}

\offprints{B. Joachimi,\\
    \email{joachimi@astro.uni-bonn.de}
}

\institute{Argelander-Institut f\"ur Astronomie (AIfA), Universit\"at Bonn, Auf dem H\"ugel 71, 53121 Bonn, Germany 
}

\date{Received 2 August 2007 / Accepted 24 October 2007}

\abstract{}
{In recent years cosmic shear, the weak gravitational lensing effect by the large-scale structure of the Universe, has proven to be one of the observational pillars on which the cosmological concordance model is founded. Several cosmic shear statistics have been developed in order to analyze data from surveys. For the covariances of the prevalent second-order measures we present simple and handy formulae, valid under the assumptions of Gaussian density fluctuations and a simple survey geometry. We also formulate these results in the context of shear tomography, i.e. the inclusion of redshift information, and generalize them to arbitrary data field geometries.}
{We define estimators for the E- and B-mode projected power spectra and show them to be unbiased in the case of Gaussianity and a simple survey geometry. From the covariance of these estimators we demonstrate how to derive covariances of arbitrary combinations of second-order cosmic shear measures. We then recalculate the power spectrum covariance for general survey geometries and examine the bias thereby introduced on the estimators for exemplary configurations.}
{Our results for the covariances are considerably simpler than and analytically shown to be equivalent to the real-space approach presented in the first paper of this series. We find good agreement with other numerical evaluations and confirm the general properties of the covariance matrices. The studies of the specific survey configurations suggest that our simplified covariances may be employed for realistic survey geometries to good approximation.}
{} 


\maketitle

\section{Introduction}
Despite the fast progress in cosmology during the past years (e.g.
\citealt{spergel07}, and references therein), still only little is
known about those 95\% of the energy content of the Universe which
is non-baryonic. Especially for the dominant dark
energy, astronomical observations provide the only source of
information to reveal more about its properties. One of the most
promising methods to shed light on the dark ingredients of the
Universe is cosmic shear, the gravitational lensing of distant
galaxies by the large-scale matter structure.

Observations of a cosmic shear signal are challenging since the shear,
i.e. the distortion of galaxy images by gravitational lensing,
is small compared to the intrinsic galaxy ellipticities, of order
1\%. Besides, measurements of galaxy ellipticities are further
hampered by atmospheric seeing, telescope optics and pixelization of
the image. Therefore, although theoretical work on light propagation
in an inhomogeneous universe exists for four decades
\citep{gunn67}, the first detection of cosmic shear was only
reported in 2000, independently by \citet{bacon00}, \citet{kaiser00}, \citet{vwaer00} and \citet{wittman00}. Recently, cosmic shear
observations have reached statistical errors on cosmological
parameters compatible to those of other established methods, see
e.g. \citet{jarvis06} with CTIO data, \citet{hoekstra06}, \citet{semboloni06} and \citet{fu07} with samples from CFHTLS, and \citet{hetterscheidt07} with an analysis of the GaBoDS survey. In combination with CMB
data, cosmic shear is able to break parameter degeneracies because it
probes the matter distribution at lower redshifts and on smaller
scales than the CMB fluctuations. Cosmic shear is also complementary to
methods such as type Ia supernovae and galaxy redshift surveys, its
strength being that the underlying theory is built on few physical
assumptions; in particular there is no dependence on the relation
between baryons and dark matter.

The statistical properties of the (line-of-sight projected) matter
distribution measured by cosmic shear are usually characterized by the
convergence power spectrum $P_\kappa(\ell)$, where $\ell$ denotes the
Fourier variable on the sky. If the matter density field is a
realization of a Gaussian random field, the second-order measure
$P_\kappa(\ell)$ fully describes its properties. This is the case if
the density perturbations are linear. 
Equivalently, other second-order statistical measures as the
correlation functions, the shear dispersion and the aperture mass (for
definitions see Sect. \ref{sec:measures}) can be used instead of
$P_\kappa(\ell)$. Of all these second-order statistical methods, the
shear correlation function is most straightforward to measure from
data, being insensitive to gaps and the field geometry of the imaged
sky region, and it provides the full second-order statistical
information contained in the data. 

In order to assess strengths and limitations of all these measures it
is important to know errors and their correlations on different scales
characterized by the covariance matrices. Under the assumption of
Gaussianity, \citet{Schneider02}, hereafter Paper I, calculated such 
covariance matrices
analytically, starting from an estimator for the correlation functions
$\xi_\pm$. By further assuming a connected survey area, negligible
boundary effects due to the finite field size and randomly placed
galaxies in the field, the ensemble average over galaxy positions of
the covariances was taken, and expressions for the covariances as a
function of the theoretical $\xi_\pm$ were derived. In this work we
aim at recalculating the covariance matrices, but now in a Fourier
space approach, leading to significantly simpler expressions. These are
then shown to be equivalent to the result of Paper I.

In Sect. \ref{sec:measures} we give an overview on second-order
measures of cosmic shear and their interrelations, using the notation
of \citet{bartelmann01}
and \citet{Schneider06}. Moreover, we briefly summarize the E- and B-mode
decomposition of the shear field. We derive estimators for E- and
B-mode power spectra in Sect. \ref{sec:estimators} and show them to be 
unbiased under the assumptions made in Paper I. The covariance
of these estimators and successively the covariances of other
second-order measures are calculated in Sect.
\ref{sec:covariances}. We demonstrate the equivalence of our formulae
to those given in Paper I in Sect.
\ref{sec:equivalence}. In addition, we evaluate the covariances
numerically and discuss our results. In Sect. \ref{sec:general} the
calculations of the foregoing sections are generalized to shear tomography,
 i.e. the inclusion of photometric redshift information, as well as to
arbitrary survey geometries. We discuss the behavior of our estimators for 
specific examples of survey geometries. Finally, we summarize our
findings and conclude in Sect. \ref{sec:conclusions}.

\section{Second-order measures of cosmic shear}
\label{sec:measures}
In order to obtain information about cosmic shear from a given data
set, galaxy ellipticities are measured at positions $\vek{ \theta}_i$
in the sky. The observed complex ellipticity $\epsilon=\epsilon_{1} +
\ic \epsilon_{2}$ contains contributions from the intrinsic source
ellipticity $\epsilon^{\rm s}$ and the shear $\gamma$ induced by
gravitational lensing, i.e. $\epsilon_i=\gamma(\vek{
\theta}_i) + \epsilon^{\rm s}_i$. By considering a pair of
galaxies at positions $\vek{\vartheta}$ and
$\vek{\vartheta}+\vek{\theta}$, one defines the tangential and cross
component of the ellipticity at position $\vek{\vartheta}$ for this
particular pair as
\eq{
\epsilon_{\rm t}=- {\rm Re} (\epsilon\; {\rm e}^{-2 \ic \varphi})\;; \hspace{1cm} 
\epsilon_\times=- {\rm Im} (\epsilon\; {\rm e}^{-2 \ic \varphi})\;,
}
where $\varphi$ is the polar angle of the separation vector
$\vek{\theta}$ (likewise for $\gamma$ and $\epsilon^{\rm s}$). Then
one defines the shear correlation functions
\eq{
\xi_\pm(\theta) := \langle \gamma_{\rm t} \gamma_{\rm t} \rangle (\theta) \pm \langle \gamma_\times \gamma_\times \rangle (\theta)\;.
}
Further second-order statistical measures are the shear dispersion $\langle |\bar{\gamma}|^2 \rangle (\theta)$, defined as the average of the mean shear in circular apertures of radius $\theta$, and the dispersion of the aperture mass in circular apertures of radius $\theta$. For an aperture located at $\theta=0$, the aperture measures are given by
\eq{
M_{\rm ap}(\theta)=\int \dd^2 \vartheta\; Q(\vartheta)\; \gamma_{\rm
  t}(\vek{\vartheta})\;; \hspace{1cm} 
M_\perp(\theta)=\int \dd^2 \vartheta\; Q(\vartheta)\; \gamma_\times(\vek{\vartheta})\;,
}
when $\vartheta=|\vek{\vartheta}|$. The weight function $Q(\vartheta)$ is axially-symmetric, but otherwise arbitrary. \citet{crittenden02} suggested a weight function with exponential fall-off. Here $Q(\vartheta)$ is chosen to be the same as in \citet[hereafter S02]{SvWM02}:
\eq{
Q(\vartheta)=\frac{6 \vartheta^2}{\pi \theta^4} \br{1-\frac{\vartheta^2}{\theta^2}} H(\theta-\vartheta)\;,
}
where $H(\theta-\vartheta)$ is the Heaviside step-function. 

The convergence or dimensionless projected surface mass density $\kappa$ is related to the shear via
\eq{
\label{eq:kapgam}
\gamma(\vek{\theta})=\frac{1}{\pi} \int \dd^2 \vartheta ~{\cal D}(\vek{\theta} - \vek{\vartheta}) \kappa(\vek{\vartheta}) ~~~\mbox{with}~~~ {\cal D}(\vek{\theta})=\frac{\theta_2^2 - \theta_1^2 - 2 \ic \theta_1 \theta_2}{|\vek{\theta}|^4}\;.
}
In ordinary lens theory $\kappa$ is a real quantity. If the complex
$\gamma$ can be calculated from $\kappa$, $\gamma_1$ and $\gamma_2$
are mutually dependent, which leads to a relation that ensures a
shear field, generated purely by lensing, to be curl-free as
demonstrated in \citet{crittenden02} and S02. As $\gamma$ is a polar quantity, 
a general shear field can be decomposed in analogy to electromagnetic
polarization into a curl-free E-mode and a divergence-free
B-mode. Although not caused by gravitational lensing, B-modes can be
present in cosmic shear data, for instance due to the underestimation
of noise or uncorrected systematic errors. Further (small) sources of
B-modes are source redshift clustering, see S02, and the
limited validity of the Born approximation, see e.g. \citet{jain00}.

For a pure lensing signal we will set the B-mode contribution to zero later on. Nevertheless we must account for noise in the B-mode channel as it is measured by the real-space estimator for the correlation functions used in Paper I, which does not discriminate between an E- or B-mode origin of galaxy ellipticities. In order to account for both E- and B-modes, we rewrite the originally purely real convergence $\kappa(\vek{\theta})$ as a complex quantity, which reads in Fourier space $\tilde{\kappa}(\vek{ \ell})=\tilde{\kappa}_{\rm E}(\vek{ \ell})+\ic \tilde{\kappa}_{\rm B}(\vek{ \ell})$. In terms of $\kappa_{\rm E}$ and $\kappa_{\rm B}$ the following power spectra are defined via
\eqa{\nn
\langle \tilde{\kappa}_{\rm E}(\vek{ \ell}) \tilde{\kappa}_{\rm E}^*(\vek{ \ell'}) \rangle&=& (2\pi)^2 \delta^{(2)}(\vek{ \ell} - \vek{\ell'}) ~P_{\rm E}(\ell)\;,\\ 
\label{eq:Pdef}
\langle \tilde{\kappa}_{\rm B}(\vek{ \ell}) \tilde{\kappa}_{\rm B}^*(\vek{ \ell'}) \rangle&=& (2\pi)^2 \delta^{(2)}(\vek{ \ell} - \vek{\ell'}) ~P_{\rm B}(\ell)\;,\\ \nn
\langle \tilde{\kappa}_{\rm E}(\vek{ \ell}) \tilde{\kappa}_{\rm B}^*(\vek{ \ell'}) \rangle&=& (2\pi)^2 \delta^{(2)}(\vek{ \ell} - \vek{\ell'}) ~P_{\rm EB}(\ell)\;,
}
where $\delta^{(2)}(\vek{\ell})$ is the two-dimensional Dirac delta-distribution. Note that all two-dimensional power spectra appearing in this work are power spectra of $\kappa$, which is why the index $\kappa$ is dropped. To lowest order, i.e. in the Born approximation and without lens-lens coupling, the E-mode power spectrum is related to the three-dimensional power spectrum $P_\delta$ of the density fluctuations via Limber's equation
\eq{
\label{eq:limber}
P_{\rm E}(\ell)=\frac{9H_0^4 \Omega_{\rm m}^2}{4 c^4} \int^{\chi_{\rm h}}_0 \dd \chi\; \frac{g^2(\chi)}{a^2(\chi)} P_\delta \br{\frac{\ell}{\chi},\chi}\;,
}
where $\chi$ is the comoving distance and $\chi_{\rm h}$ the comoving horizon distance. Hence, $P(\ell)$ is the projection of $P_\delta$ along the line-of-sight, weighted with the lensing efficiency $g(\chi)=\int^{\chi_h}_{\chi} \dd \chi' p(\chi') (1-\chi/\chi')$, where $p(\chi)$ is the distribution of source galaxies in distance. For simplicity we have assumed a spatially flat geometry. From now on we set $P_{\rm EB}(\ell) \equiv 0$ because the cross power is expected to vanish for a statistically parity-invariant shear field, as is the case for its corresponding correlation function $\xi_\times(\theta)=\langle \gamma_{\rm t} \gamma_\times \rangle (\theta)$.

The Fourier transform of (\ref{eq:kapgam}) reads
\eq{
\label{eq:gamkap}
\tilde{\gamma}(\vek{ \ell})={\rm e}^{2 \ic \beta} \tilde{\kappa}(\vek{ \ell})\;,
}
where $\beta$ is the polar angle of $\vek{\ell}$, and where $\kappa$ is now a complex quantity. By means of this relation and (\ref{eq:Pdef}) one can derive the second-order shear measures as a function of the E- and B-mode power spectra, see e.g. S02,
\eqa{\nn
\xi_+(\theta)%&=&\langle \gamma(0) \gamma^*(\vek{ \theta}) \rangle\\ \nn
%&=& \int \frac{\dd^2 \ell}{(2 \pi)^2} \int \frac{\dd^2 \ell'}{(2 \pi)^2} {\rm e}^{\ic \svek{ \ell'} \cdot \svek{\theta}} \langle \gamma(\vek{ \ell}) \gamma^*(\vek{ \ell'}) \rangle\\ 
&=& \int^\infty_0 \frac{\dd \ell ~\ell}{2 \pi} ~J_0(\ell \theta) \{ P_{\rm E}(\ell) + P_{\rm B}(\ell) \}\;; \hspace{2cm}
\xi_-(\theta) = \int^\infty_0 \frac{\dd \ell ~\ell}{2 \pi} ~J_4(\ell \theta) \{ P_{\rm E}(\ell) - P_{\rm B}(\ell) \}\;,\\ 
\label{eq:filters}
\langle M_{\rm ap}^2 \rangle(\theta) &=& \int^\infty_0 \frac{\dd \ell ~\ell}{2 \pi} ~\frac{576 J_4^2(\ell \theta)}{(\ell \theta)^4} P_{\rm E}(\ell)\;; \hspace{2.3cm}
\langle M_\perp^2 \rangle(\theta) = \int^\infty_0 \frac{\dd \ell ~\ell}{2 \pi} ~\frac{576 J_4^2(\ell \theta)}{(\ell \theta)^4} P_{\rm B}(\ell)\;,\\ \nn
\ba{ |\bar{\gamma}|^2 } (\theta) &=& \int^\infty_0 \frac{\dd \ell ~\ell}{2 \pi} ~\frac{4 J_1^2(\ell \theta)}{(\ell \theta)^2} \bc{P_{\rm E}(\ell) + P_{\rm B}(\ell)}\;,
}
where $J_\mu(x)$ is the Bessel function of the first kind of order $\mu$.

So the above measures are all linear functionals of the power spectra, each employing a different filter. As a consequence they are interrelated, as shown by \citet{crittenden02}. Thus, in practice only one measure has to be estimated from the data, the others can then be deduced. As real data contains gaps, masks and CCD defects, the placing of apertures needed for $\langle |\bar{\gamma}|^2 \rangle (\theta)$ and $\langle M_{\rm ap}^2 \rangle(\theta)$ is impractical, so that the correlation functions are the preferred primary observables. Due to the broad filter especially of $\xi_+$ the signal is strong, but mixes information about the power spectrum over a wide range of scales. On the contrary, the aperture mass statistics employ a narrow filter, which can be replaced by a Dirac $\delta$-distribution with an error of only 10\% \citep{bartelmann99}. Thus the signal is small, but probes the power spectrum very locally, and measures on different scales quickly decorrelate (see Sect. \ref{sec:equivalence}). In addition to this, (\ref{eq:filters}) shows that $\langle M_{\rm ap}^2 \rangle(\theta)$ is sensitive only to E-modes, whereas $\langle M_\perp^2 \rangle(\theta)$, being sensitive only to B-modes, can be used to quantify systematic errors etc. The filters of the second-order measures also determine the form of their covariance matrices, as will be discussed in Sect. \ref{sec:equivalence}.

\section{Power spectrum estimators}
\label{sec:estimators}
Consider now $N$ galaxies at positions $\vek{\theta}_i$ with measured ellipticities $\epsilon_i=\epsilon_{1,i}+\ic \epsilon_{2,i}$ in a data field covering a solid angle $A$. Repeating a calculation by \citet{kaiser98}, we write
\eq{
\label{eq:kaiser98}
\tilde{\epsilon}_\alpha(\vek{ \ell}) = \sum_{i=1}^N \epsilon_{\alpha,i}\; {\rm e}^{\ic \svek{ \ell} \cdot \svek{ \theta_i}}
= \int \dd^2 \theta\; {\rm e}^{\ic \svek{ \ell} \cdot \svek{ \theta}} n(\vek{ \theta}) \gamma_\alpha(\vek{ \theta}) + \sum_i \epsilon_{\alpha,i}^{\rm s}\; {\rm e}^{\ic \svek{ \ell} \cdot \svek{ \theta}_i}\;,
}
where in the second step $\epsilon_{\alpha,i}$ was split up into shear and intrinsic ellipticity. Note that $\tilde{\epsilon}_\alpha(\vek{ \ell})$ is the discrete Fourier transform of the real ellipticity component $\epsilon_\alpha$ so that $\tilde{\epsilon}^*_\alpha(\vek{ \ell})=\tilde{\epsilon}_\alpha(-\vek{ \ell})$ holds. The introduction of $n(\vek{ \theta}):=\sum_i \delta^{(2)}(\vek{\theta}  - \vek{\theta_i})$ regains a continuous Fourier transform for the shear. Transforming $\gamma_\alpha(\vek{ \theta})$ back to Fourier space then yields
\eq{
\label{eq:epsgeneral}
\tilde{\epsilon}_\alpha(\vek{ \ell}) = \int \frac{\dd^2 \ell'}{(2\pi)^2} \tilde{\gamma}_\alpha(\vek{ \ell'}) \tilde{n}(\vek{ \ell} - \vek{\ell'}) + \sum_i \epsilon_{\alpha,i}^{\rm s}\; {\rm e}^{\ic \svek{ \ell} \cdot \svek{ \theta}_i}
\hspace{.5cm} \mbox{with} \hspace{.5cm}
\tilde{n}(\vek{\ell})=\int \dd^2 \theta\; {\rm e}^{\ic \svek{ \ell} \cdot \svek{ \theta}} n(\vek{ \theta})\;. 
}
The explicit dependence of $n(\vek{ \theta})$ on the galaxy positions hampers analytical progress. To simplify this expression, we make the assumption that the galaxies in the field are uniformly sampled, which means that $n(\vek{ \theta})$ can be replaced by its ensemble average over all galaxy positions 
\eqnn{
E(n(\vek{ \theta}))=\bar{n}\; \Pi(\vek{\theta}) 
\hspace{.5cm} \mbox{with} \hspace{.5cm}
E := \prod^N_{j=1} \br{\frac{1}{A} \int \dd^2 \theta_j}\;,
}
where $\bar{n}=N/A$ is the mean number density of galaxies in the field and $\Pi(\vek{ \theta})$ is the aperture function, which is $1$ for $\vek{\theta}$ within the field and $0$ else. In particular, $A=\int \dd^2 \theta ~\Pi(\vek{ \theta})$. With the definition
\eq{
 \Delta(\vek{\ell}) := \frac{\tilde{n}(\vek{\ell})}{(2\pi)^2 \bar{n}} = \int \frac{\dd^2 \theta}{(2 \pi)^2}\; {\rm e}^{\ic \svek{ \ell} \cdot \svek{ \theta}}\; \Pi(\vek{\theta})
}
one can now write $\tilde{n}(\vek{\ell})=(2\pi)^2\bar{n}\Delta(\vek{\ell})$. Note that $(2\pi)^2\Delta(\vek{\ell})$ is the Fourier transform of the aperture. If one further assumes that the observed field is simply connected and that all relevant angles are considerably smaller than the extent of the field, i.e. $|\vek{\theta}|^2 \ll A$, then the aperture function $\Pi$ can be approximated by unity, so that consequently $\Delta(\vek{\ell}) \approx \delta^{(2)}(\vek{\ell})$. Through this (\ref{eq:epsgeneral}) takes on the form
\eq{
\label{eq:epsapprox}
\tilde{\epsilon}_\alpha(\vek{ \ell}) \approx \bar{n} \tilde{\gamma}_\alpha(\vek{\ell}) + \sum_i \epsilon_{\alpha,i}^{\rm s}\; {\rm e}^{\ic \svek{ \ell} \cdot \svek{ \theta}_i}\;.
}
In the following calculations terms containing ellipticities to the square will appear. In this case the approximation for $\Delta(\vek{\ell})$ is executed as follows
\eq{
\label{eq:deltaapprox}
|\Delta(\vek{\ell})|^2 \approx \delta^{(2)}(\vek{\ell})\; \Delta(\vek{\ell}) = \delta^{(2)}(\vek{\ell})\; \Delta(0) =  \delta^{(2)}(\vek{\ell})\; \frac{A}{(2\pi)^2}\;.
}
With these considerations at hand we now define the following estimators for the E- and B-mode power spectra,
\eqa{\nn
\hat{P}_{\rm E}(\bar{\ell}):=\frac{1}{\bar{n}^2 A A_R(\bar{\ell})} \int_{A_R(\bar{\ell})} \dd^2 \ell ~\left| ~~\tilde{\epsilon}_1(\vek{ \ell }) \cos 2\beta + \tilde{\epsilon}_2(\vek{ \ell }) \sin 2\beta \right|^2 -\frac{\sigma_\epsilon^2}{2\bar{n}}\;,\\
\label{eq:estimators} 
\hat{P}_{\rm B}(\bar{\ell}):=\frac{1}{\bar{n}^2 A A_R(\bar{\ell})} \int_{A_R(\bar{\ell})} \dd^2 \ell ~\left| -\tilde{\epsilon}_1(\vek{ \ell }) \sin 2\beta + \tilde{\epsilon}_2(\vek{ \ell }) \cos 2\beta \right|^2 -\frac{\sigma_\epsilon^2}{2\bar{n}}\;.
}
These estimators are band powers for disjunct bins $\bar{\ell}$ in Fourier space, constructed by averaging over annuli with mean radius $\bar{\ell}$ and area $A_R(\bar{\ell})=2 \pi \bar{\ell} \Delta \ell$. $\sigma_\epsilon^2$ denotes the dispersion of the complex source ellipticities. The Fourier ellipticities $\tilde{\epsilon}_\alpha(\vek{ \ell })$ in (\ref{eq:estimators}) are calculated from the data via (\ref{eq:kaiser98}). In the following we demonstrate that these estimators are unbiased under the assumptions made above.

On calculating the expectation value, one splits up the image ellipticities into a shear and a source ellipticity part following (\ref{eq:epsapprox}). In order to process the latter terms, consider the definition of the source ellipticity dispersion $\langle \epsilon^{\rm s}_i \epsilon^{{\rm s}*}_j \rangle=\delta_{ij}\sigma_\epsilon^2$. Moreover, as the Universe is isotropic, we expect the combination $\langle \epsilon^{\rm s}_i \epsilon^{{\rm s}}_j \rangle$, which has a net orientation, to vanish. From these two complex correlators one concludes for the ellipticity components
\eq{
\label{eq:eps}
\langle \epsilon_{\alpha,i}^{\rm s} \epsilon_{\beta,j}^{\rm s} \rangle = \delta_{ij} \delta_{\alpha\beta} \frac{\sigma_\epsilon^2}{2}\;; \hspace{.5cm} \alpha,\beta=\{1,2\}\;.
}
The source ellipticity terms are processed via this relation and result in $\sigma_\epsilon^2/2\bar{n}$, so that they cancel with the last term in (\ref{eq:estimators}). In case of E-modes, the remaining shear terms yield
\eq{
\ba{\hat{P}_{\rm E}(\bar{\ell})} = \frac{1}{A A_R(\bar{\ell})} \int_{A_R(\bar{\ell})} \dd^2 \ell \ba{ |\tilde{\gamma}_1(\vek{ \ell })|^2 \cos^2 2\beta+ |\tilde{\gamma}_2(\vek{ \ell })|^2 \sin^2 2\beta + 2 \bc{\tilde{\gamma}_1(\vek{ \ell }) \tilde{\gamma}^*_2(\vek{ \ell }) + \tilde{\gamma}^*_1(\vek{ \ell }) \tilde{\gamma}_2(\vek{ \ell })} \sin 2\beta \cos 2\beta }\;.
}
The shears are replaced by means of the expressions
\eq{
\label{eq:kappagamma}
\tilde{\kappa}_{\rm E}(\vek{ \ell}) = \tilde{\gamma}_1(\vek{ \ell}) \cos 2\beta + \tilde{\gamma}_2(\vek{ \ell}) \sin 2\beta\;;
\hspace{1cm}
\tilde{\kappa}_{\rm B}(\vek{ \ell}) = -\tilde{\gamma}_1 (\vek{ \ell}) \sin 2\beta + \tilde{\gamma}_2(\vek{ \ell}) \cos 2\beta\;,
}
derived from (\ref{eq:gamkap}). Via (\ref{eq:Pdef}) one then inserts the power spectrum. Formally, at this stage a Dirac delta-distribution with argument $0$ appears, but as terms quadratic in the ellipticities are involved, the approximation (\ref{eq:deltaapprox}) applies, leading to
\eq{
\ba{\hat{P}_{\rm E}(\bar{\ell})} =  \frac{1}{A A_R(\bar{\ell})} \int_{A_R(\bar{\ell})} \dd^2 \ell ~(2\pi)^2 \Delta(0) P_{\rm E}(\ell) = \frac{1}{A_R(\bar{\ell})} \int_{A_R(\bar{\ell})} \dd^2 \ell ~P_{\rm E}(\ell) = P_{\rm E}(\bar{\ell})\;.
}
In complete analogy, one can show $\hat{P}_{\rm B}$ to be an unbiased estimator of the power spectrum, too. In Sect. \ref{sec:general} we will drop the assumptions on $\Delta(\vek{\ell})$ made here and examine how well the estimators (\ref{eq:estimators}) are suited for more realistic situations.

\section{Covariances}
\label{sec:covariances}
Next we are going to calculate the covariance of the estimators defined above. Introducing the shorthand notation $\Delta P_{\rm X}(\ell) \equiv \hat{P}_{\rm X}(\ell) - P_{\rm X}(\ell)$, the covariance reads
\eq{
{\rm Cov}(P_{\rm X}; \bar{\ell},\bar{\ell'}) := \langle \Delta P_{\rm X}(\bar{\ell}) \Delta P_{\rm X}(\bar{\ell'}) \rangle = \langle \hat{P}_{\rm X}(\bar{\ell}) \hat{P}_{\rm X}(\bar{\ell'}) \rangle - P_{\rm X}(\bar{\ell}) P_{\rm X}(\bar{\ell'})\;.
}
For simplicity we keep to the case of the E-mode power spectrum in the following; the B-mode calculation is analogous. Plugging in the estimators yields for the correlator
\eqa{
\label{eq:correlatorPE}
\langle \hat{P}_{\rm E}(\bar{\ell}) \hat{P}_{\rm E}(\bar{\ell'}) \rangle &=&  \int_{A_R(\bar{\ell})} \frac{\dd^2 \ell}{\bar{n}^2 A A_R(\bar{\ell})} \int_{A_R(\bar{\ell'})} \frac{\dd^2 \ell'}{\bar{n}^2 A A_R(\bar{\ell'})}
\ba{\bigl|\tilde{\epsilon}_1(\vek{\ell}) \cos 2\beta + \tilde{\epsilon}_2(\vek{\ell}) \sin 2\beta \bigr|^2 \bs{\tilde{\epsilon}_1(\vek{\ell'}) \cos 2\beta' + \tilde{\epsilon}_2(\vek{\ell'}) \sin 2\beta'}^2}\\ \nn
&-& \frac{\sigma_\epsilon^2}{2\bar{n}} \int_{A_R(\bar{\ell})} \frac{\dd^2 \ell}{\bar{n}^2 A A_R(\bar{\ell})} \ba{\bigl|\tilde{\epsilon}_1(\vek{\ell}) \cos 2\beta + \tilde{\epsilon}_2(\vek{\ell}) \sin 2\beta \bigr|^2}
- \frac{\sigma_\epsilon^2}{2\bar{n}} \int_{A_R(\bar{\ell'})} \frac{\dd^2 \ell'}{\bar{n}^2 A A_R(\bar{\ell'})} \ba{\bs{\tilde{\epsilon}_1(\vek{\ell'}) \cos 2\beta' + \tilde{\epsilon}_2(\vek{\ell'}) \sin 2\beta'}^2} + \frac{\sigma_\epsilon^4}{4\bar{n}^2}.
}
By comparison with (\ref{eq:estimators}) the integrals of the second and third term amount to $P_{\rm E}(\bar{\ell})+\sigma_\epsilon^2/2\bar{n}$ and $P_{\rm E}(\bar{\ell'})+\sigma_\epsilon^2/2\bar{n}$, respectively. The expansion of the first term results in a number of four-point correlators of ellipticities. If the shear field is assumed to be Gaussian, which is realistic if one considers scales sufficiently large that effects of non-linear density evolution do not have an effect on the shear field, this can be written as a sum of products of two-point correlators. The four-point correlators of the intrinsic source ellipticities can be decomposed analogously. So terms of type $\langle \tilde{\epsilon}_\alpha(\vek{ \ell}) \tilde{\epsilon}_\beta(\vek{ \ell'}) \rangle$ remain, which can be further processed via (\ref{eq:epsapprox}) as follows:
\eq{
\langle \tilde{\epsilon}_\alpha(\vek{ \ell}) \tilde{\epsilon}^*_\beta(\vek{ \ell'}) \rangle = \bar{n}^2 \langle \tilde{\gamma}_\alpha(\vek{ \ell}) \tilde{\gamma}^*_\beta(\vek{ \ell'}) \rangle + \sum_{ij} \langle \epsilon_{\alpha,i}^{\rm s} \epsilon_{\beta,j}^{\rm s} \rangle {\rm e}^{\ic \br{\svek{\ell} \cdot \svek{\theta}_i - \svek{\ell'} \cdot \svek{\theta}_j}} 
= \bar{n}^2 \langle \tilde{\gamma}_\alpha(\vek{ \ell}) \tilde{\gamma}^*_\beta(\vek{ \ell'}) \rangle + \delta_{\alpha\beta} \frac{\sigma_\epsilon^2}{2} \br{\sum_{i} {\rm e}^{\ic \br{\svek{\ell}  - \svek{\ell'}} \cdot \svek{\theta}_i} }\;, 
}
where for the first step we assumed that intrinsic ellipticities and shear are not correlated, while in the second step we used (\ref{eq:eps}). The term in brackets can be recognized as $\tilde{n}(\vek{\ell} - \vek{\ell'})$, whereby we can approximate the equation as done before, leading to
\eq{
\label{eq:correpsapprox}
\langle \tilde{\epsilon}_\alpha(\vek{ \ell}) \tilde{\epsilon}^*_\beta(\vek{ \ell'}) \rangle = \bar{n}^2 \langle \tilde{\gamma}_\alpha(\vek{ \ell}) \tilde{\gamma}^*_\beta(\vek{ \ell'}) \rangle + \delta_{\alpha\beta} \frac{\sigma_\epsilon^2}{2} (2\pi)^2 \bar{n} ~\delta^{(2)}(\vek{\ell} - \vek{\ell'})\;.
}
Correlators of the form $\langle \tilde{\epsilon}_\alpha(\vek{ \ell}) \tilde{\epsilon}_\beta(\vek{ \ell'}) \rangle$ and $\langle \tilde{\epsilon}^*_\alpha(\vek{ \ell}) \tilde{\epsilon}^*_\beta(\vek{ \ell'}) \rangle$ can be dealt with analogously by employing $\tilde{\epsilon}^*_\alpha(\vek{ \ell})=\tilde{\epsilon}_\alpha(-\vek{ \ell})$.

By expressing the correlators $\langle \tilde{\gamma}_\alpha(\vek{ \ell}) \tilde{\gamma}^*_\beta(\vek{ \ell'}) \rangle$ in terms of correlators $\langle \tilde{\kappa}_{\rm X}(\vek{\ell}) \tilde{\kappa}_{\rm Y}^*(\vek{\ell'}) \rangle$ (for X,Y$=\{E,B\}$), using (\ref{eq:kappagamma}), (\ref{eq:correlatorPE}) can be written in the form
\eq{\label{eq:correlatorPE2}
\langle \hat{P}_{\rm E}(\bar{\ell}) \hat{P}_{\rm E}(\bar{\ell'}) \rangle = \int_{A_R(\bar{\ell})} \frac{\dd^2 \ell}{\bar{n}^2 A A_R(\bar{\ell})} \int_{A_R(\bar{\ell'})} \frac{\dd^2 \ell'}{\bar{n}^2 A A_R(\bar{\ell'})} \bc{{\cal A}_{\rm E} + {\cal B}_{\rm E} + {\cal C}_{\rm E} }
- \frac{\sigma_\epsilon^2}{2\bar{n}} \br{P_{\rm E}(\bar{\ell})+P_{\rm E}(\bar{\ell'})} - \frac{\sigma_\epsilon^4}{4\bar{n}^2}\;,
}
with
\eqa{\nn
{\cal A}_{\rm E} &=& \bar{n}^4 \br{\langle \tilde{\kappa}_{\rm E}(\vek{\ell}) \tilde{\kappa}_{\rm E}^*(\vek{\ell}) \rangle \langle \tilde{\kappa}_{\rm E}(\vek{\ell'}) \tilde{\kappa}_{\rm E}^*(\vek{\ell'}) \rangle + 2 \langle \tilde{\kappa}_{\rm E}(\vek{\ell}) \tilde{\kappa}_{\rm E}^*(\vek{\ell'}) \rangle^2 }
= \bar{n}^4 \br{A^2 P_{\rm E}(\ell) P_{\rm E}(\ell') + 2 ~\delta^{(2)}(\vek{\ell} - \vek{\ell'}) ~A (2\pi)^2 P^2_{\rm E}(\ell)}\;,\\ \nn
{\cal B}_{\rm E} &=& \frac{\sigma_\epsilon^2}{2} (2\pi)^2 \bar{n}^3 \br{\frac{A}{(2\pi)^2} \langle \tilde{\kappa}_{\rm E}(\vek{\ell}) \tilde{\kappa}_{\rm E}^*(\vek{\ell}) \rangle + \frac{A}{(2\pi)^2} \langle \tilde{\kappa}_{\rm E}(\vek{\ell'}) \tilde{\kappa}_{\rm E}^*(\vek{\ell'}) \rangle + 4 ~\delta^{(2)}(\vek{\ell} - \vek{\ell'}) \langle \tilde{\kappa}_{\rm E}(\vek{\ell}) \tilde{\kappa}_{\rm E}^*(\vek{\ell}) \rangle }\\ \nn
&=& \frac{\sigma_\epsilon^2}{2} (2\pi)^2 \bar{n}^3 \br{\frac{A^2}{(2\pi)^2} P_{\rm E}(\ell) + \frac{A^2}{(2\pi)^2} P_{\rm E}(\ell') + 4 ~\delta^{(2)}(\vek{\ell} - \vek{\ell'}) ~A\; P_{\rm E}(\ell)}\;,\\ \nn
{\cal C}_{\rm E} &=& \frac{\sigma_\epsilon^4}{4} (2\pi)^4 \bar{n}^2 \br{ \frac{A^2}{(2\pi)^4} + 2 ~\delta^{(2)}(\vek{\ell} - \vek{\ell'}) \frac{A}{(2\pi)^2} }\;,
}
where in the second steps, respectively, we have made use of (\ref{eq:Pdef}) together with (\ref{eq:deltaapprox}). Here and in the following, ${\cal A}$ denotes cosmic variance terms, ${\cal C}$ is the shot noise contribution proportional to $\sigma_\epsilon^4$ and ${\cal B}$ stands for the mixed term. Shot noise is caused by the intrinsic ellipticity dispersion of the source galaxies and usually dominates on small angular scales (in the Gaussian approximation). The fact that a data field of finite extent is used to estimate statistical measures leads to another noise component called cosmic variance, which prevails on scales larger than about 5$'$ or $\ell \apprle 2000$, respectively, see e.g. \citet{hu01}.

By subtracting the product of the mean of the estimators, (\ref{eq:correlatorPE2}) turns into
\eq{
\langle \Delta P_{\rm E}(\bar{\ell}) \Delta P_{\rm E}(\bar{\ell'}) \rangle = 2 \cdot (2\pi)^2 A \bar{n}^2 \int_{A_R(\bar{\ell})} \frac{\dd^2 \ell}{\bar{n}^2 A A_R(\bar{\ell})} \int_{A_R(\bar{\ell'})} \frac{\dd^2 \ell'}{\bar{n}^2 A A_R(\bar{\ell'})} \delta^{(2)}(\vek{\ell} - \vek{\ell'}) \br{\bar{n}\; P_{\rm E}(\ell) + \frac{\sigma_\epsilon^2}{2}}^2.
}
Note that the Dirac delta-distribution can only have a non-vanishing result if the two integration areas $A_R(\bar{\ell})$ and $A_R(\bar{\ell'})$ overlap, i.e. if $\bar{\ell} = \bar{\ell'}$. Performing the integrations and repeating the above considerations in the B-mode case, one arrives at
\eq{
\label{eq:covP1}
\langle \Delta P_{\rm X}(\bar{\ell}) \Delta P_{\rm X}(\bar{\ell'}) \rangle = \frac{4 \pi}{A \bar{\ell} \Delta \ell} \left( P_{\rm X}(\bar{\ell}) + \frac{\sigma_\epsilon^2}{2\bar{n}} \right)^2 \delta_{\bar{\ell}\bar{\ell'}} \;\; \mbox{for} \;\; X=\{E,B\}\;,
}
where the area of an annulus with thickness $\Delta \ell$, $A_R(\bar{\ell})=2 \pi \bar{\ell} \Delta \ell$, has already been inserted. Equation (\ref{eq:covP1}) for E-modes is in agreement with the results presented in \citet{kaiser98}. A similar calculation yields
\eq{
\langle \Delta P_{\rm E}(\bar{\ell}) \Delta P_{\rm B}(\bar{\ell'}) \rangle = 0\;,
}
i.e. as expected the E- and B-mode power spectra are not correlated.

By means of (\ref{eq:filters}) the covariances of all possible combinations of second-order cosmic shear measures can now easily be obtained. For instance, one gets for the covariance of the correlation function $\xi_+$
\eqa{\nn
\langle \Delta \xi_+ (\theta_1) ~\Delta \xi_+ (\theta_2) \rangle
&\approx& \frac{1}{4 \pi^2} \sum_{\bar{\ell},\bar{\ell'}}  \bar{\ell} ~\bar{\ell'} {\Delta \ell}^2 J_0(\bar{\ell} \theta_1) J_0(\bar{\ell'} \theta_2) \left\langle \left( \Delta P_{\rm E}(\bar{\ell}) + \Delta P_{\rm B}(\bar{\ell}) \right) \left( \Delta P_{\rm E}(\bar{\ell'}) + \Delta P_{\rm B}(\bar{\ell'}) \right) \right\rangle \\
&=& \frac{1}{4 \pi^2} \sum_{\bar{\ell},\bar{\ell'}}  \bar{\ell} ~\bar{\ell'} {\Delta \ell}^2 J_0(\bar{\ell} \theta_1) J_0(\bar{\ell'} \theta_2)  \left( \langle \Delta P_{\rm E}(\bar{\ell}) \Delta P_{\rm E}(\bar{\ell'}) \rangle + \langle  \Delta P_{\rm B}(\bar{\ell}) \Delta P_{\rm B}(\bar{\ell'}) \rangle \right)\;,
}
where we have used a discretized version of the transformation because the power spectrum covariance is given for $\ell$-bins. After insertion of (\ref{eq:covP1}) the covariance depends linearly on $\Delta \ell$; therefore, considering the limit $\Delta \ell \rightarrow 0$, the sum transforms back into an integral,
\eq{
\label{eq:xicov0}
 \langle \Delta \xi_+ (\theta_1) ~\Delta \xi_+ (\theta_2) \rangle
= \frac{1}{\pi A} \int^\infty_0 \dd \ell ~\ell ~J_0(\ell \theta_1) J_0(\ell \theta_2) \left\{ \left( P_{\rm E}(\ell) + \frac{\sigma_\epsilon^2}{2\bar{n}} \right)^2 + \left(  P_{\rm B}(\ell) + \frac{\sigma_\epsilon^2}{2\bar{n}} \right)^2 \right\}\;.
}
Similarly the results for other combinations of second-order measures are derived, so that for the case of correlation functions one arrives at
\eqa{
\label{eq:xicov1}
 \langle \Delta \xi_- (\theta_1) ~\Delta \xi_- (\theta_2) \rangle
&=& \frac{1}{\pi A} \int^\infty_0 \dd \ell ~\ell ~J_4(\ell \theta_1) J_4(\ell \theta_2) \left\{ \left( P_{\rm E}(\ell) + \frac{\sigma_\epsilon^2}{2\bar{n}} \right)^2 + \left(  P_{\rm B}(\ell) + \frac{\sigma_\epsilon^2}{2\bar{n}} \right)^2 \right\}\;,\\
\label{eq:xicov2}
 \langle \Delta \xi_+ (\theta_1) ~\Delta \xi_- (\theta_2) \rangle
&=& \frac{1}{\pi A} \int^\infty_0 \dd \ell ~\ell ~J_0(\ell \theta_1) J_4(\ell \theta_2) \left\{ \left( P_{\rm E}(\ell) + \frac{\sigma_\epsilon^2}{2\bar{n}} \right)^2 - \left(  P_{\rm B}(\ell) + \frac{\sigma_\epsilon^2}{2\bar{n}} \right)^2 \right\}\;,
}
whereas for the aperture mass we get
\eq{
\label{eq:mapcov}
\ba{\Delta \langle M_{\rm ap}^2 \rangle(\theta_1) ~\Delta \langle M_{\rm ap}^2 \rangle(\theta_2)} = \frac{576^2}{\pi A \theta_1^4 \theta_2^4} \int^\infty_0 \dd \ell ~\ell^{-7} ~J_4^2(\ell \theta_1) J_4^2(\ell \theta_2) \br{ P_{\rm E}(\ell) + \frac{\sigma_\epsilon^2}{2\bar{n}} }^2.
}
Up to now the covariances are continuous functions of $\theta$ although in reality angular scales are binned. If the binning is sufficiently small and the integrands in the above equations are smooth functions of $\theta_1$ and $\theta_2$, one can simply replace the angles by the central values of the bins. However, the shot noise term of (\ref{eq:xicov0}) and (\ref{eq:xicov1}) is not smooth, but can be further processed by means of the orthogonality relation of the Bessel functions
\eq{
\label{eq:orthogonality}
\int^\infty_0 \dd \ell ~\ell ~J_\mu(\ell \theta_1) J_\mu(\ell \theta_2) = \frac{1}{\theta_1} ~\delta(\theta_1 - \theta_2)\;.
}
The transition to a discrete set of angular bins is done by the relation between Kronecker and Dirac $\delta$'s: $\delta_{\theta_1\theta_2} \Delta\theta^{-1} \rightarrow  \delta^{(2)}(\theta_2-\theta_1)$ for $\Delta\theta \rightarrow 0$. Introducing the number of galaxy pairs in an annulus of bin radius $\theta$ and thickness $\Delta \theta$, $N_{\rm p}(\theta)=2\pi\, \theta\, \Delta \theta\, A\, \bar{n}^2$, one can write (\ref{eq:xicov0}) and (\ref{eq:xicov1}) in the directly applicable form
\eq{
\label{eq:xicovbin}
\langle \Delta \xi_\pm (\theta_1) ~\Delta \xi_\pm (\theta_2) \rangle
= \frac{1}{\pi A} \int^\infty_0 \dd \ell ~\ell ~J_{0/4}(\ell \theta_1) J_{0/4}(\ell \theta_2) \bc{ P^2_{\rm E}(\ell) + P^2_{\rm B}(\ell) + \frac{\sigma_\epsilon^2}{\bar{n}} \br{P_{\rm E}(\ell) + P_{\rm B}(\ell)} } + \frac{\sigma_\epsilon^4}{N_{\rm p}(\theta_1)} ~\delta_{\theta_1 \theta_2}\;.
}
The shot noise term in (\ref{eq:xicov2}) cancels, so that in this case no further processing is needed.

\section{Equivalence to real space covariances}
\label{sec:equivalence}
In Paper I covariances of the correlation functions and other second-order measures were obtained by starting directly with an estimator for $\xi_\pm$ and by taking an ensemble average over all galaxy positions of the resulting covariances. These calculations were based on the same assumptions as this work; therefore our results and those from Paper I are equivalent, which is proven in the following.

\subsection{Correlation function covariances}
Starting from (\ref{eq:xicovbin}), the covariance for $\xi_+$ can again be split up into cosmic variance, shot noise and mixed term as follows 
\eqa{\nn
{\cal A_{+\,+}} &=& \frac{1}{\pi A} \int^\infty_0 \dd \ell ~\ell ~\bc{P_{\rm E}^2(\ell) + P_{\rm B}^2(\ell)} J_0(\ell \theta_1) J_0(\ell \theta_2)\;,\\ 
\label{eq:C++terms}
{\cal B_{+\,+}} &=& \frac{\sigma_\epsilon^2}{\pi A \bar{n}} \int^\infty_0 \dd \ell ~\ell ~\bc{P_{\rm E}(\ell) + P_{\rm B}(\ell)} J_0(\ell \theta_1) J_0(\ell \theta_2)\;,\\ \nn
{\cal C_{+\,+}} &=& \frac{\sigma_\epsilon^4}{N_{\rm p}(\theta_1)} ~\delta_{\theta_1 \theta_2}\;,
}
where here and in the following, we drop the arguments $\theta_1$, $\theta_2$ of the partial correlators ${\cal A_{+\,+}}$ etc. The quantity ${\cal C_{+\,+}}$ is already in the desired form of Paper I, (27). Inverting the first two equations of (\ref{eq:filters}) by means of the orthogonality relation (\ref{eq:orthogonality}) results in (see also S02)
\eq{
\label{eq:Pxi}
P_{\rm E}(\ell) = \pi \int^\infty_0 \dd \theta ~\theta \bc{\xi_+(\theta) J_0(\ell \theta) + \xi_-(\theta) J_4(\ell \theta)}\;; \hspace{1cm}
P_{\rm B}(\ell) = \pi \int^\infty_0 \dd \theta ~\theta \bc{\xi_+(\theta) J_0(\ell \theta) - \xi_-(\theta) J_4(\ell \theta)}\;,
}
from which one concludes $P_{\rm E}(\ell)+P_{\rm B}(\ell)=\int \dd^2 \theta  ~\xi_+(\theta) {\rm e}^{\ic \svek{\ell} \cdot \svek{\theta}}$. If one inserts this expression and the definition of the Bessel function in the form $J_0(x)=\int^{2\pi}_0 \frac{\dd \varphi}{2\pi} {\rm e}^{\pm \ic x \cos \varphi}$ into (\ref{eq:C++terms}), the mixed term yields
\eq{
\label{eq:mixedxipl}
{\cal B_{+\,+}} = \frac{\sigma_\epsilon^2}{8\pi^4 A \bar{n}} \int \dd^2 \ell \int \dd^2 \theta \int^{2\pi}_0 \dd \varphi_1 \int^{2\pi}_0 \dd \varphi_2 ~\xi_+(\theta)\; {\rm e}^{\ic \svek{\ell} \cdot (\svek{\theta} - \svek{\theta}_1 + \svek{\theta}_2) }
= \frac{2 \sigma_\epsilon^2}{\pi A \bar{n}} \int^\pi_0 \dd \varphi ~\xi_+\left( \sqrt{\theta_1^2 + \theta_2^2 - 2\theta_1\theta_2\cos \varphi} \right)\;.
}
After integrating over $\vek{\ell}$ and $\vek{\theta}$ the argument of $\xi_+$ depends only on the difference $\varphi=\varphi_2-\varphi_1$ of the polar angles $\varphi_1$, $\varphi_2$ of $\vek{\theta}_1$, $\vek{\theta}_2$, respectively, so that one of the $\varphi$-integrals can be carried out. The last equality in (\ref{eq:mixedxipl}) is now equivalent to (32) in Paper I.

Due to the free choice of the sign of the exponential in the Bessel functions $J_0$ and $J_4$ in this and the following calculations the results should be invariant with respect to the signs of the angles in the exponent, in this case $\vek{\theta}_1$ and $\vek{\theta}_2$. This is indeed the case as can be shown by redefinition of the relevant integration variables. We choose the signs of the Bessel function exponentials such that the results of Paper I are reproduced in their original form.

In order to derive the cosmic variance term, we make use of (\ref{eq:Pxi}) to write 
\eq{
\label{eq:splitP}
P_{\rm E}^2(\ell)+P_{\rm B}^2(\ell) = \frac{1}{2} \int \dd^2 \theta \int \dd^2 \theta' \br{\xi_+(\theta) \xi_+(\theta') + \xi_-(\theta) \xi_-(\theta') {\rm e}^{4\ic (\varphi_\theta - \varphi_{\theta'})} } {\rm e}^{\ic \svek{\ell} \cdot (\svek{\theta} - \svek{\theta'})}\;,
}
which then yields, after writing the Bessel functions in exponential form,
\eqa{\nn
{\cal A_{+\,+}}&=& \frac{1}{(2\pi)^4 A} \int \dd^2 \ell \int \dd^2 \theta \int \dd^2 \theta' \int^{2\pi}_0 \dd \varphi_1 \int^{2\pi}_0 \dd \varphi_2 ~{\rm e}^{\ic \svek{\ell} \cdot (\svek{\theta} + \svek{\theta}_1 - \svek{\theta'} + \svek{\theta}_2)}
 \br{ \xi_+(\theta) \xi_+(\theta') + \xi_-(\theta) \xi_-(\theta') {\rm e}^{4\ic (\varphi_\theta - \varphi_{\theta'})}}\\ 
\label{eq:cvxipl}
&=& \frac{1}{(2\pi)^2 A} \int \dd^2 \phi \int^{2\pi}_0 \dd \varphi_1 \int^{2\pi}_0 \dd \varphi_2 \br{ \xi_+(\psi_{\rm a}) \xi_+(\psi_{\rm b}) + \xi_-(\psi_{\rm a}) \xi_-(\psi_{\rm b}) {\rm e}^{4\ic (\varphi_{\rm a} - \varphi_{\rm b})}}\\ \nn
&=& \frac{2}{\pi A} \int^\infty_0 \dd \phi ~\phi \int^{\pi}_0 \dd \varphi_1 \int^{\pi}_0 \dd \varphi_2 ~\xi_+(\psi_{\rm a}) \xi_+(\psi_{\rm b})
+ \frac{1}{2\pi A} \int^\infty_0 \dd \phi ~\phi \int^{2\pi}_0 \dd \varphi_1 \int^{2\pi}_0 \dd \varphi_2 ~\xi_-(\psi_{\rm a}) \xi_-(\psi_{\rm b}) \cos 4(\varphi_{\rm a}-\varphi_{\rm b})\;, 
}
where we have integrated over $\vek{\ell}$ and $\vek{\theta'}$ in order to arrive at the second equality and successively defined the vectors $\vek{\phi}:=\vek{\theta}+\vek{\theta}_1$, $\vek{\psi}_{\rm a}:=\vek{\phi}-\vek{\theta}_1 \equiv \vek{\theta}$ and $\vek{\psi}_{\rm b}:=\vek{\phi}+\vek{\theta}_2$ to adjust our notation to that of Paper I. The quantities $\varphi_{\rm a}$ and $\varphi_{\rm b}$ denote the polar angles of $\vek{\psi}_{\rm a}$ and $\vek{\psi}_{\rm b}$, respectively. It is easily verified that (\ref{eq:cvxipl}) is identical to (34) of Paper I, so that the equivalence of $\langle \Delta \xi_+ (\theta_1) ~\Delta \xi_+ (\theta_2) \rangle$ is proven.

In complete analogy we repeat the calculations for $\langle \Delta \xi_- (\theta_1) ~\Delta \xi_- (\theta_2) \rangle$ and $\langle \Delta \xi_+ (\theta_1) ~\Delta \xi_- (\theta_2) \rangle$. Note that ${\cal C_{-\,-}} \equiv {\cal C_{+\,+}}$ and ${\cal C_{+\,-}} \equiv 0$. The mixed terms then read
\eq{
{\cal B_{-\,-}} = \frac{\sigma_\epsilon^2}{8\pi^4 A \bar{n}} \int \dd^2 \ell \int \dd^2 \theta \int^{2\pi}_0 \dd \varphi_1 \int^{2\pi}_0 \dd \varphi_2 ~\xi_+(\theta) {\rm e}^{\ic \svek{\ell} \cdot (\svek{\theta} - \svek{\theta}_1 + \svek{\theta}_2) } {\rm e}^{4\ic (\varphi_2 - \varphi_1)}
= \frac{2 \sigma_\epsilon^2}{\pi A \bar{n}} \int^\pi_0 \dd \varphi ~\xi_+\left( \sqrt{\theta_1^2 + \theta_2^2 - 2\theta_1\theta_2\cos \varphi} \right) \cos 4\varphi
}
and
\eqa{
{\cal B_{+\,-}}
&=& \frac{\sigma_\epsilon^2}{8\pi^4 A \bar{n}} \int \dd^2 \ell \int \dd^2 \theta \int^{2\pi}_0 \dd \varphi_1 \int^{2\pi}_0 \dd \varphi_2 ~\xi_-(\theta) {\rm e}^{\ic \svek{\ell} \cdot (\svek{\theta} + \svek{\theta}_1 - \svek{\theta}_2) } {\rm e}^{4\ic (\varphi_\theta - \varphi_2)}\\ \nn
&=& \frac{2 \sigma_\epsilon^2}{\pi A \bar{n}} \int^\pi_0 \dd \varphi ~\xi_- \left( \sqrt{\theta_1^2 + \theta_2^2 - 2\theta_1\theta_2\cos \varphi} \right)
 \bc{\sum^4_{j=0} {4 \choose j} (-1)^j ~\theta_1^j ~\theta_2^{4-j} \cos(j\varphi)} \br{\theta_1^2 + \theta_2^2 - 2 \theta_1 \theta_2 \cos \varphi}^{-2}\;.
}
In the last step we have made use of a geometrical identity based on the cosine theorem, $\cos(\zeta - \varphi_2)=\bc{\theta_1 \cos(\varphi_2-\varphi_1) - \theta_2}/\br{|\vek{\theta}_2 - \vek{\theta}_1|}$, where $\zeta$ denotes the polar angle of $\vek{\theta}_2 - \vek{\theta}_1$. When replacing $P_{\rm E}^2(\ell) + P_{\rm B}^2(\ell)$ again as in (\ref{eq:splitP}), we find
\eqa{\nn
{\cal A_{-\,-}}
&=& \frac{1}{2\pi A} \int^\infty_0 \dd \phi ~\phi \int^{2\pi}_0 \dd \varphi_1 \int^{2\pi}_0 \dd \varphi_2 ~\xi_-(\psi_{\rm a}) \xi_-(\psi_{\rm b}) \cos 4(\varphi_1+\varphi_2-\varphi_{\rm a}-\varphi_{\rm b})\\
&+& \frac{1}{2\pi A} \int^\infty_0 \dd \phi ~\phi \int^{2\pi}_0 \dd \varphi_1 \int^{2\pi}_0 \dd \varphi_2 ~\xi_+(\psi_{\rm a}) \xi_+(\psi_{\rm b}) \cos 4(\varphi_1-\varphi_2)\;. 
}
Finally, now writing
\eq{
P_{\rm E}^2(\ell)-P_{\rm B}^2(\ell) = \frac{1}{2} \int \dd^2 \theta \int \dd^2 \theta' \br{\xi_+(\theta) \xi_-(\theta') {\rm e}^{-4\ic \varphi'} + \xi_-(\theta) \xi_+(\theta') {\rm e}^{4\ic \varphi} } {\rm e}^{\ic \svek{\ell} \cdot (\svek{\theta} - \svek{\theta'})}\;,
}
again derived from (\ref{eq:Pxi}), we get
\eqa{
{\cal A_{+\,-}}
&=& \frac{1}{\pi A} \int^\infty_0 \dd \phi ~\phi \int^{2\pi}_0 \dd \varphi_1 \int^{2\pi}_0 \dd \varphi_2 ~\xi_+(\psi_{\rm a}) \xi_-(\psi_{\rm b}) \cos 4(\varphi_2-\varphi_{\rm b})\;.
}
Hence, the results for all correlation function covariances are equivalent to those of the real-space approach of Paper I.

\begin{figure}[t]
\includegraphics[scale=.678,angle=270]{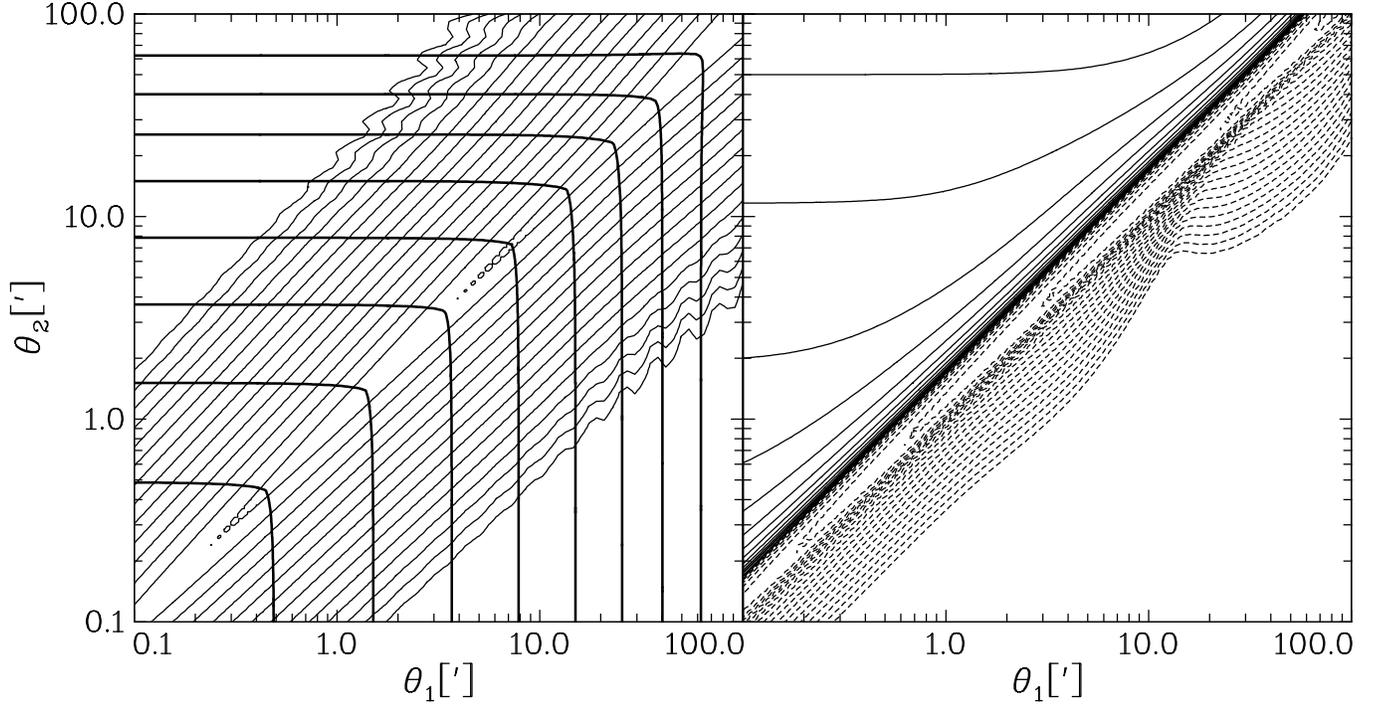}
\caption{Covariance of the correlation functions. \textit{Left panel}: Contour plots for $\langle \Delta \xi_+ (\theta_1) ~\Delta \xi_+ (\theta_2) \rangle$ (thick curves) and $\langle \Delta \xi_- (\theta_1) ~\Delta \xi_- (\theta_2) \rangle$, each with shot noise term removed, which would only yield a bin-size dependent contribution to the diagonal. For $\langle \Delta \xi_+ (\theta_1) ~\Delta \xi_+ (\theta_2) \rangle$ contours are linearly spaced, starting at $10^{-9}$ in the top right corner and with highest value of $10^{-8}$. For $\langle \Delta \xi_- (\theta_1) ~\Delta \xi_- (\theta_2) \rangle$ contours are logarithmically spaced, differing by factors of $2.25$ each with a maximum value of $10^{-10}$ in the top right corner. \textit{Right panel}: Contour plot for $\langle \Delta \xi_+ (\theta_1) ~\Delta \xi_- (\theta_2) \rangle$. Contours are logarithmically spaced, differing by factors of $1.5$ each. Solid contours correspond to positive values with maximum at $4.4\times10^{-10}$ in the top left corner, dashed ones to negative values with minimum at $-1.5\times10^{-10}$ in the top right corner.}
\label{fig:covcorr}
\end{figure}

Moreover, this equivalence can be verified numerically, too. The explicit evaluation of (\ref{eq:xicov0}) to (\ref{eq:mapcov}) is straightforward as it merely consists of a one-dimensional, though highly oscillatory integral, in contrast to the more complicated formulae derived in Paper I. In order to allow for direct comparisons, we make use of the same cosmological model as Paper I, i.e. a $\Lambda$CDM universe with $\Omega_{\rm m}=0.3$ and $\Omega_{\rm \Lambda}=0.7$. The power spectrum of density fluctuations is characterized by the primordial slope $n_{\rm s}=1$, the shape parameter $\Gamma=0.21$ and the normalization $\sigma_8=1.0$. For the computation of the linear power spectrum the fit formula of \citet{bardeen86} is used, together with the description of the non-linear evolution by \citet{PeacockDodds}. To calculate the projected power spectrum, a galaxy redshift distribution $p(z) \propto z^2 \exp \bc{-(z/z_0)^\beta}$ with $z_0=1.0$ and $\beta=1.5$ is assumed. Finally, the survey properties are specified by a galaxy number density $\bar{n}=30 ~{\rm arcmin}^{-2}$ and an intrinsic galaxy ellipticity dispersion of $\sigma_\epsilon=0.3$. Since all covariances scale with the field size as $A^{-1}$, we have chosen a fiducial size of $A=1\,{\rm deg}^2$ for Figs. \ref{fig:covcorr} and \ref{fig:mapcov}.

In Fig. \ref{fig:covcorr} the covariances of the correlation functions with the shot noise term removed are plotted. We can affirm the general characteristics of the covariance matrices; however, there are deviations from the results of Paper I exceeding numerical uncertainties. For $\langle \Delta \xi_+ (\theta_1) ~\Delta \xi_+ (\theta_2) \rangle$ we find an offset of the order $10\%$ which is nearly constant in the plotted region, whereas $\langle \Delta \xi_- (\theta_1) ~\Delta \xi_- (\theta_2) \rangle$ quickly tends to zero with increasing separation from the diagonal instead of decreasing further to negative values. Deviations are largest for the lower right corner of the $\langle \Delta \xi_+ (\theta_1) ~\Delta \xi_- (\theta_2) \rangle$-plane where we obtain negative values close to the diagonal and values close to zero off the diagonal. These differences can be traced back to the fact that in Paper I the correlation functions used as input are approximated by power laws for angles larger than $5\,$deg, which has a non-negligible effect on the results due to the infinite range of integration over angles. Moreover, we find very good agreement between the three covariances of Fig. \ref{fig:covcorr} and those given in \citet{kilbinger04}, Paper II of this series, who numerically calculated the covariances from the formulae given in Paper I without taking the ensemble average over galaxy positions.

The form of the covariance matrices can be explained in terms of the filter functions introduced in (\ref{eq:filters}). $J_0(x)$ and $J_4(x)$ behave similarly for large $x$, but in contrast to $J_4(x)$, which tends to 0 for small $x$, $J_0(x)$ remains positive and is therefore a much broader filter. As a consequence, $\xi_+(\theta')$ is strongly correlated with all $\xi_+$ at scales $\theta < \theta'$, which leads to the nearly square contours in Fig. \ref{fig:covcorr}. The narrow filter $J_4(x)$ results in a quick decorrelation of $\xi_-$ for increasingly different angular scales and thus in a diagonal structure of the covariance matrix.

\subsection{Aperture mass covariance}

\begin{figure}[t]
\begin{minipage}[c]{0.55\textwidth}
\centering
\includegraphics[scale=.68,angle=270]{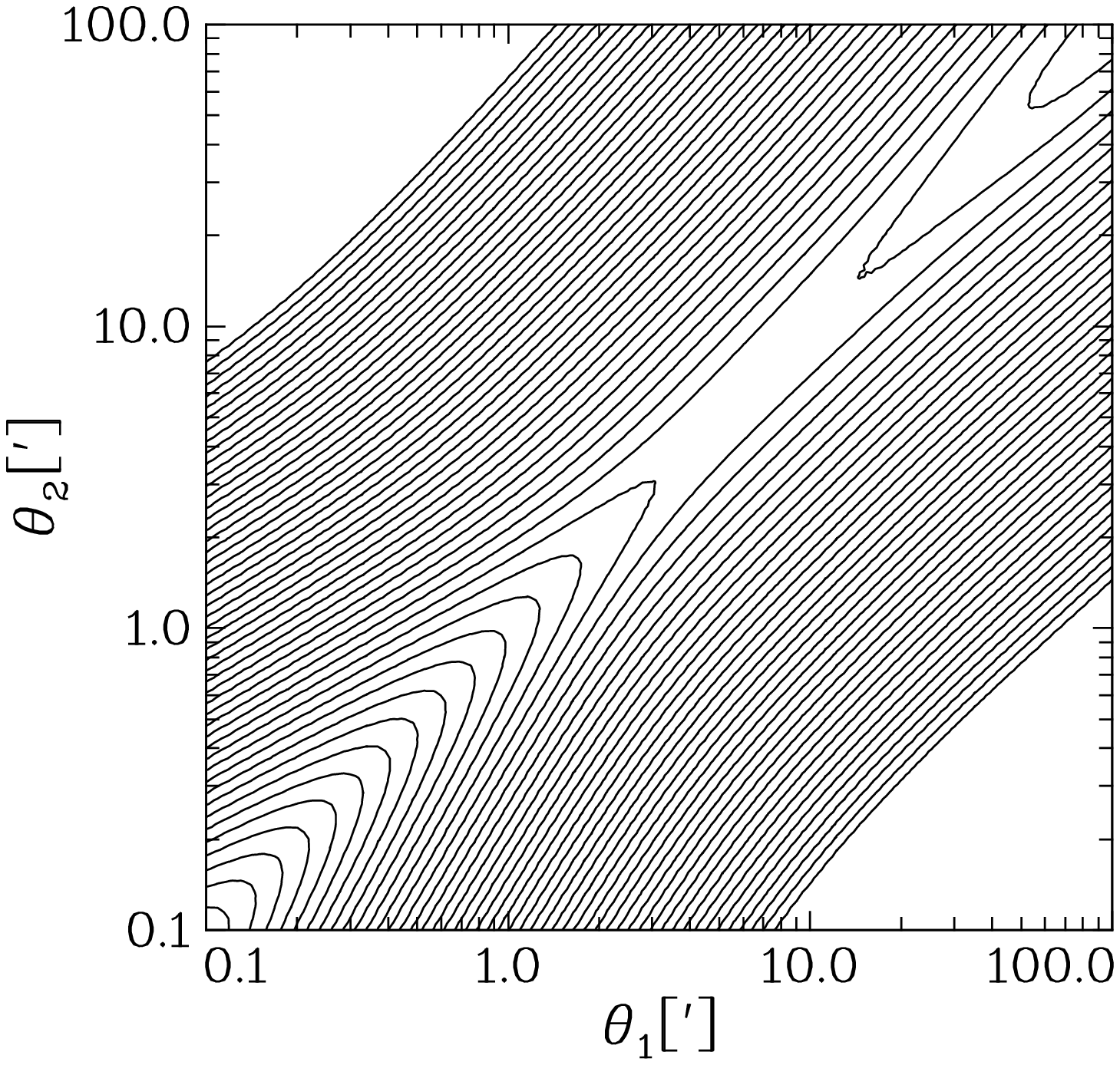}
\end{minipage}% 
\begin{minipage}[c]{0.45\textwidth}
\centering
\includegraphics[scale=.68]{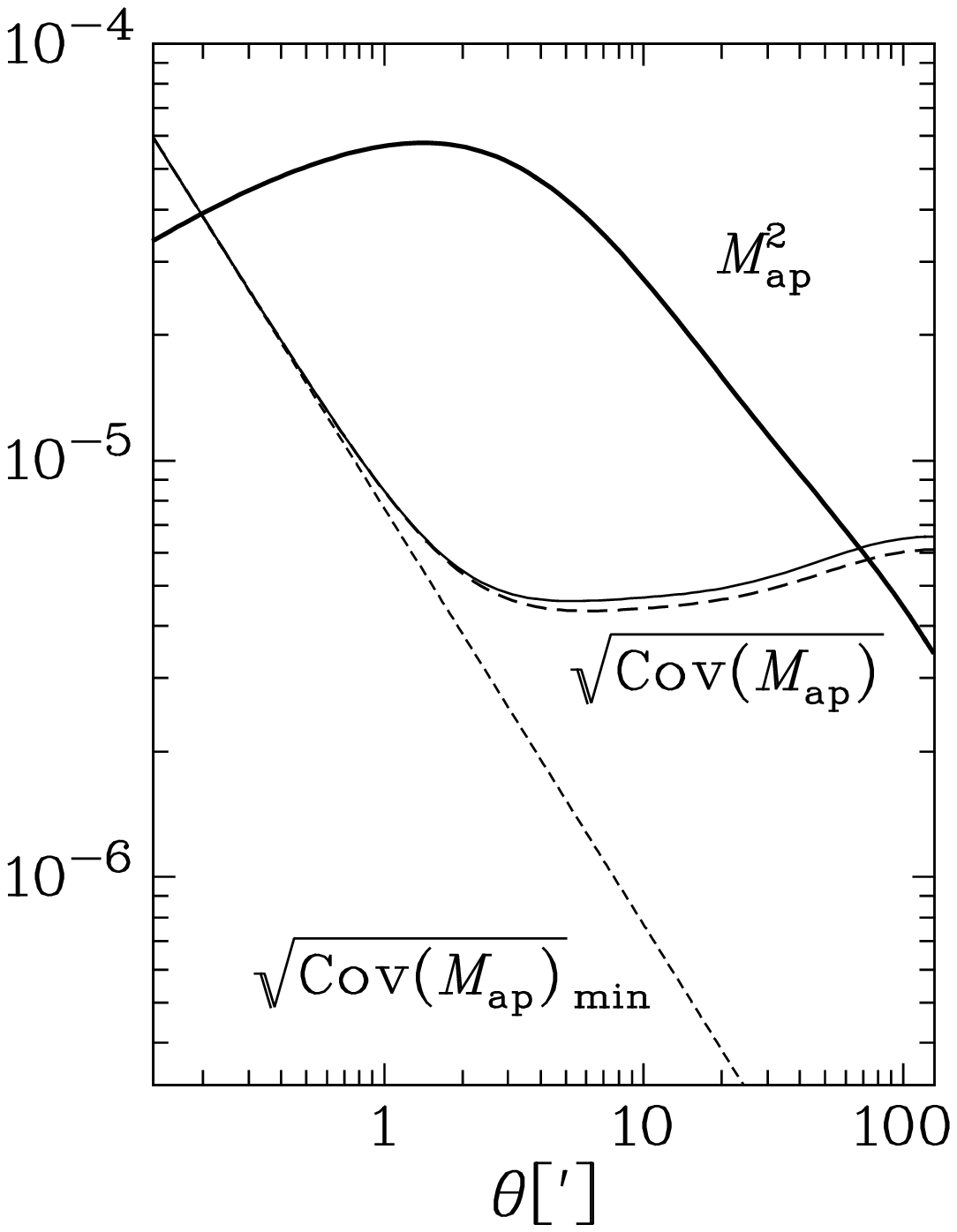}
\end{minipage}
\caption{Covariance of $M^2_{\rm ap}$. \textit{Left panel}: Contour plot for the covariance of $M^2_{\rm ap}$. Contours are logarithmically spaced, differing by factors of $1.5$ each with highest value of $4.4\times10^{-8}$ in the lower left corner. \textit{Right panel}: The thin solid curve depicts the square-root of the auto-variance of $M^2_{\rm ap}$. In addition, the same measure, calculated from the scaled expression in \citet{Schneider98}, see (\ref{eq:mapS98}), is displayed as thick long-dashed curve. The analytic expression for the minimum auto-variance (dashed line) in absence of cosmic variance is plotted, too. $\langle M^2_{\rm ap} \rangle (\theta)$ is shown as thick solid line for comparison.}
\label{fig:mapcov}
\end{figure} 

Paper I also gave the covariance of the aperture mass for an estimator ${\cal M}(\theta)$ of $\langle M_{\rm ap}^2 \rangle$, which is a function of the correlation function estimators and reads
\eq{
\label{eq:mapestimator}
{\cal M}(\theta) = \frac{\Delta \vartheta}{2\theta^2} \sum_{i=1}^{2m} \vartheta_i \bc{ \hat{\xi}_+(\vartheta_i) T_+\br{\frac{\vartheta_i}{\theta}} + \hat{\xi}_-(\vartheta_i) T_-\br{\frac{\vartheta_i}{\theta}} }\;,
}
where $\Delta \vartheta$ is the angular bin width, the $\vartheta_i$ are the centers of these bins, and furthermore $\theta=m \Delta \vartheta$ with $m$ being an integer. The functions $T_+$ and $T_-$ are given in S02. This form of the estimator for $\langle M_{\rm ap}^2 \rangle$ is insensitive to B-modes. Since the covariance of the aperture mass in Paper I is given in terms of the correlation function covariances, one can directly plug in the results of the previous section and show the equivalence to (\ref{eq:mapcov}). However, the transformation from $\xi_\pm$ to $\langle M_{\rm ap}^2 \rangle$ is merely based on (\ref{eq:filters}), so that the equivalence proven in the foregoing section readily implies the equivalence also for the aperture mass covariance.

In absence of gravitational lensing the shot noise term ${\cal C_{\rm ap}}$ yields the only contribution and is therefore the minimum covariance. Either from (\ref{eq:mapcov}) or from Paper I, (42) one obtains
\eq{
{\cal C_{\rm ap}}  
%\frac{\sigma_\epsilon^4}{8 \pi A \bar{n}^2} \frac{576^2}{\theta_1^2 \theta_2^2} \int_0^\infty \dd \vartheta ~\vartheta \int_0^\infty \frac{\dd t}{t^3} \int_0^\infty \frac{\dd t'}{t'^3} ~J_4^2(t) J_4^2(t') \bc{J_0\br{\frac{\vartheta}{\theta_1}t} J_0\br{\frac{\vartheta}{\theta_2}t'} + J_4\br{\frac{\vartheta}{\theta_1}t} J_4\br{\frac{\vartheta}{\theta_2}t'}}\\
\label{eq:mapC}
= \frac{\sigma_\epsilon^4}{4 \pi A \bar{n}^2} \frac{576^2}{\theta_1^4 \theta_2^4} \int^\infty_0 \dd \ell ~\ell^{-7} ~J_4^2(\ell \theta_1) J_4^2(\ell \theta_2)\;.
}
Note that in the absence of B-modes ${\cal C_{\rm ap}}$ constitutes the full covariance matrix of $\langle M_\perp^2 \rangle$. If one considers the auto-variance, the integral in (\ref{eq:mapC}) can be evaluated analytically giving
\eq{
\ba{{\Delta \langle M_{\rm ap}^2 \rangle}^2 (\theta)}_{\rm min} = \frac{1327104}{1819125 \pi^3} ~\frac{\sigma_\epsilon^4}{A \bar{n}^2 \theta^2}\;,
}
in agreement with the numerical evaluation of Paper I.

Figure \ref{fig:mapcov} displays the numerical results for the covariance matrix of $\langle M^2_{\rm ap} \rangle$ when using the same survey geometry and cosmological parameters as before. We choose again a fiducial field size of $A=1\,{\rm deg}^2$. Due to the very narrow filter function of $\langle M^2_{\rm ap} \rangle$, the correlation is strongly concentrated towards the diagonal. When the ratio of angular scales exceeds 2, the covariance has decreased to less than a third. The structure of the diagonal is analyzed further in the right panel where the square-root of the auto-variance is shown. On small scales the noise of intrinsic galaxy ellipticities dominates (at least in the Gaussian approximation), whereas on large scales cosmic variance takes over. The overall functional form is identical to the results of Paper I, the differences again stemming from their approximation of correlation functions as discussed in the foregoing section.

Moreover, we compare our result for the auto-variance with the one derived in \citet{Schneider98} who started with a direct estimator for $\langle M^2_{\rm ap} \rangle$. In the Gaussian case the approximate solution therein reads
\eq{
\label{eq:mapS98}
\ba{{\Delta \langle M_{\rm ap}^2 \rangle}^2 (\theta)}=\frac{1}{N_{\rm f}} \br{\sqrt{2}\ba{M_{\rm ap}^2} + \frac{\sigma_\epsilon^2 G}{\sqrt{2} \pi \bar{n} \theta^2}}^2,
}
where $G$ is a factor of order unity depending on the choice of the filter function $Q$ for $M_{\rm ap}$; for our choice, $G=6/5$. $N_{\rm f}$ is the number of statistically independent apertures which can be placed on the survey area. We set $1/N_{\rm f}:=\eta\, \theta^2/A$, where $\eta$ is a fudge factor determined from the arbitrarily chosen condition that the shot-noise terms of both approaches should have the same value. The resulting auto-variance is also plotted in Fig. \ref{fig:mapcov}, being identical to the power spectrum approach within a few per cent. The actual noise of the $\langle M^2_{\rm ap} \rangle$-estimator will be larger though, because $\eta \sim 0.6$ is smaller than unity. This is expected because for the estimators (\ref{eq:estimators}) every ellipticity in the field is used, whereas in placing circular apertures information is left out even if one allows the apertures to partly overlap.

\section{Generalizations}
\label{sec:general}

\subsection{Shear tomography}

The inclusion of photometric redshift information about the distance of the observed galaxies allows one to measure the effects of gravitational lensing of the large-scale structure in three dimensions, enabling the determination of cosmological parameters with considerably tighter constraints, e.g. \citet{hu99,bacon04}. In particular, it is possible to follow the evolution of structure with time, which depends on the density of dark energy and its equation of state \citep{hu03,benabed04}. Besides, redshift information is essential to eliminate systematic effects such as intrinsic alignment of galaxies from the cosmic shear signal, see e.g. \citet{king03,hirata04}. Therefore it is of interest to review the results obtained for the power spectrum covariance in the context of shear tomography. Note that we retain our assumptions used in Sects. \ref{sec:estimators} to \ref{sec:equivalence}.

\begin{figure}[t]
\begin{minipage}[c]{0.2\textwidth}
\centering
\caption{Covariance matrix of tomographic shear correlation functions $\xi_+^{(ij)}$, plotted without shot noise term. Each block consists of the covariance matrix of the correlation functions given on the margin. Within a block, entries correspond to 65 logarithmically spaced $\theta$-bins in the range between 2$'$ and 40$'$. Contours are logarithmically spaced with the lowest values of about $10^{-10}$ in the upper left corner (darkest shading) and largest values of $2\times10^{-8}$ in the lower right corner (brightest shading).}
\label{fig:tomography}
\end{minipage}% 
\begin{minipage}[c]{0.8\textwidth}
\centering
\includegraphics[scale=.75,angle=270]{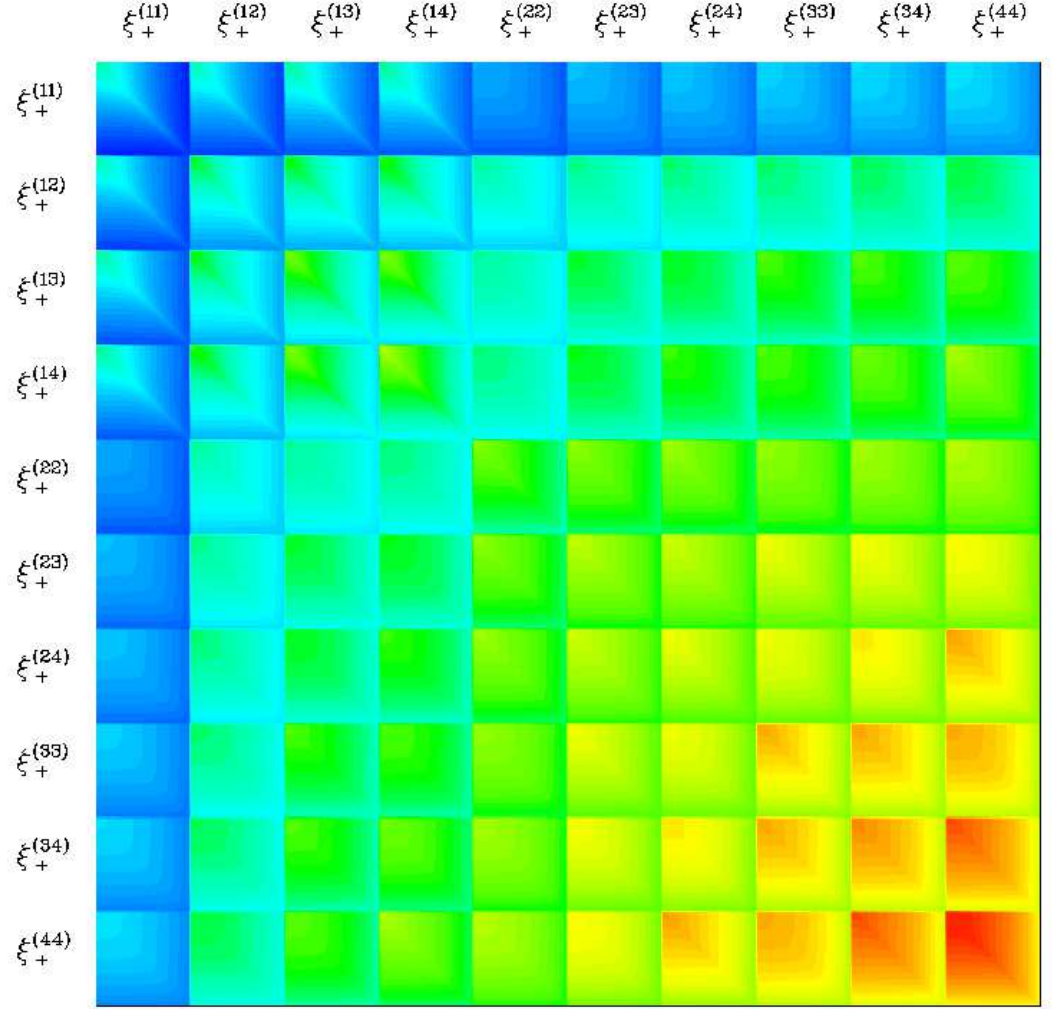}
\end{minipage}
\end{figure}

Consider $N_{\rm bin}$ redshift bins into which all galaxies can be sorted. For each bin $k$ an equivalent lens plane with convergence $\kappa^{(k)}_{\rm E}$ as well as ellipticities $\epsilon^{(k)}$ and shears $\gamma^{(k)}$ are defined. Then one can construct auto- and cross-correlation tomography power spectra
\eqa{\nn
\ba{ \tilde{\kappa}^{(k)}_{\rm E}(\vek{ \ell}) \tilde{\kappa}^{(l)*}_{\rm E}(\vek{ \ell'}) } &=& (2\pi)^2 \delta^{(2)}(\vek{ \ell} - \vek{\ell'}) ~P^{(kl)}_{\rm E}(\ell)\;,\\ 
\label{eq:Pdeftom}
\ba{ \tilde{\kappa}^{(k)}_{\rm B}(\vek{ \ell}) \tilde{\kappa}^{(l)*}_{\rm B}(\vek{ \ell'}) } &=& (2\pi)^2 \delta^{(2)}(\vek{ \ell} - \vek{\ell'}) ~P^{(kl)}_{\rm B}(\ell)\;,\\ \nn
\ba{ \tilde{\kappa}^{(k)}_{\rm E}(\vek{ \ell}) \tilde{\kappa}^{(l)*}_{\rm B}(\vek{ \ell'}) } &=& 0\;,
}
where we again assumed the E- and B-mode cross power spectra to vanish due to parity symmetry, i.e. $P^{(kl)}_{\rm EB}(\ell) \equiv 0$. The tomography power spectra are connected to the three-dimensional matter power spectrum via (\ref{eq:limber}) with modified lensing efficiencies $g^{(k)}(\chi)=\int^{\chi_{\rm h}}_\chi \dd \chi' p^{(k)}(\chi') (1-\chi/\chi')$ for bin $k$. Here $p^{(k)}(\chi)$ stands for the distribution of galaxies within bin $k$. The distributions of different bins will in general overlap, for instance due to uncertainties in the determination of photometric redshifts. From the power spectra one obtains second-order shear tomography measures and their interrelations, as for instance
\eq{
\xi^{(kl)}_\pm(\theta) = \ba{ \gamma^{(k)}_{\rm t}
\gamma^{(l)}_{\rm t}  } (\theta) \pm \ba{ \gamma^{(k)}_\times \gamma^{(l)}_\times } (\theta)
\label{eq:xitom}
= \int^\infty_0 \frac{\dd \ell ~\ell}{2 \pi} ~J_{0/4}(\ell \theta) \bc{ P^{(kl)}_{\rm E}(\ell) \pm P^{(kl)}_{\rm B}(\ell) }\;.
}
Let the number of galaxy pairs in bin $k$ be $N^{(k)}$, and the corresponding number density $\bar{n}^{(k)}=N^{(k)}/A$. Then (\ref{eq:correpsapprox}) is generalized to
\eq{
\ba{ \tilde{\epsilon}^{(k)}_\alpha(\vek{ \ell}) \tilde{\epsilon}^{(l)*}_\beta(\vek{ \ell'}) } = \bar{n}^{(k)} \bar{n}^{(l)} \ba{ \tilde{\gamma}^{(k)}_\alpha(\vek{ \ell}) \tilde{\gamma}^{(l)*}_\beta(\vek{ \ell'}) } + \delta_{\alpha\beta} \delta_{kl} \frac{\sigma_\epsilon^2}{2} (2\pi)^2 \bar{n}^{(k)} ~\delta^{(2)}(\vek{\ell} - \vek{\ell'})\;.
}
The $\delta_{kl}$ in the second term appears because intrinsic ellipticities of galaxies are only correlated with themselves, so that there cannot be a correlation between different bins. Now we define an estimator for the shear tomography power spectrum
\eq{
\hat{P}^{(kl)}_{\rm E}(\bar{\ell}):=\frac{1}{\bar{n}^{(k)} \bar{n}^{(l)}  A A_R(\bar{\ell})} \int_{A_R} \dd^2 \ell ~\br{\tilde{\epsilon}^{(k)}_1(\vek{ \ell }) \cos 2\beta + \tilde{\epsilon}^{(k)}_2(\vek{ \ell }) \sin 2\beta }
\br{\tilde{\epsilon}^{(l)*}_1(\vek{ \ell }) \cos 2\beta + \tilde{\epsilon}^{(l)*}_2(\vek{ \ell }) \sin 2\beta } - \delta_{kl} \frac{\sigma_\epsilon^2}{2\bar{n}^{(k)}}\;,
}
and likewise for $\hat{P}^{(kl)}_{\rm B}(\bar{\ell})$. Both estimators are unbiased, i.e. $\ba{ \hat{P}^{(kl)}_{\rm X}(\bar{\ell}) } = P^{(kl)}_{\rm X}(\bar{\ell})$ for $X=\{E,B\}$ and all $k,l$. Repeating the calculation as in Sect. \ref{sec:covariances}, one arrives at the covariances of the power spectra, which read
\eq{
\label{eq:covPtom}
\ba{ \Delta P^{(ij)}_{\rm X}(\bar{\ell}) \Delta P^{(kl)}_{\rm X}(\bar{\ell'}) } = \frac{2\pi}{A \bar{\ell} \Delta \ell} \br{ \bar{P}^{(ik)}_{\rm X}(\bar{\ell})\bar{P}^{(jl)}_{\rm X}(\bar{\ell}) + \bar{P}^{(il)}_{\rm X}(\bar{\ell})\bar{P}^{(jk)}_{\rm X}(\bar{\ell}) }  \delta_{\bar{\ell}\bar{\ell'}}
\;\;\; \mbox{with} \;\; \bar{P}^{(kl)}_{\rm X}:=P^{(kl)}_{\rm X}+\delta_{kl} \frac{\sigma_\epsilon^2}{2\bar{n}^{(k)}}\;.
}
This result is in agreement with the covariance derived for instance in \citet{takada04}, \citet{ma05} and \citet{huterer06}. Note that as in the non-tomographic case, any E- and B-mode cross-covariances of the tomography power spectra vanish. By means of the second equality in (\ref{eq:xitom}) and corresponding relations, the covariances of shear tomography second-order measures can now easily be obtained. We confine ourselves to the example of $\xi_+$, for which one gets in the case of $P_{\rm B}(\ell)\equiv0$
\eq{
 \ba{ \Delta \xi^{(ij)}_+ (\theta_1) ~\Delta \xi^{(kl)}_+ (\theta_2) }
= \frac{1}{2 \pi A} \int^\infty_0 \dd \ell ~\ell ~J_0(\ell \theta_1) J_0(\ell \theta_2) 
\bc{ \br{ \bar{P}^{(ik)}_{\rm E}(\ell)\bar{P}^{(jl)}_{\rm E}(\ell) + \bar{P}^{(il)}_{\rm E}(\ell)\bar{P}^{(jk)}_{\rm E}(\ell) } + \frac{\sigma_\epsilon^4}{4\bar{n}^{(i)}\bar{n}^{(j)}} \br{\delta_{ik}\delta_{jl}+\delta_{il}\delta_{jk}}}\;.
}
\citet{simon04} calculated such a covariance matrix from mock galaxy catalogs drawn from Gaussian random fields by means of a Monte Carlo simulation. As a practical example we recalculate this matrix, the result given in Fig. \ref{fig:tomography}. We employ the same cosmological model and redshift distribution as \citet{simon04}, which are identical to the ones used in Sect. \ref{sec:equivalence}. Similarly we adopt a mean galaxy density of $\bar{n}=30\,{\rm arcmin}^{-2}$, an intrinsic ellipticity dispersion of $\sigma_\epsilon=0.3$, a single square field of observation of size $1.25\times1.25\,{\rm deg}^2$, and four equally sized disjunct redshift bins between 0 and 3. Angular scales range from 2$'$ to 40$'$, divided into 65 bins. Again we leave out the shot noise term in the plot as it would only contribute to the diagonal of each sub-matrix.

The overall behavior of values in Fig. \ref{fig:tomography} is the same as in \citet{simon04}, the apparent differences in the structure within the blocks being due to the linear spacing of the $\theta$-bins in \citet{simon04}. As expected, the structure of blocks of type $\ba{ \Delta \xi^{(ii)}_+ ~\Delta \xi^{(ii)}_+ }$ is similar to Fig. \ref{fig:covcorr}. The large values for the covariance of the correlation functions at the highest redshifts are caused by both increasing $P^{(ii)}_{\rm E}(\ell)$ and the low number of galaxies in the highest redshift bin. 

\subsection{Arbitrary survey geometry}

Data of real cosmic shear surveys will in general not satisfy our assumption of a simple, filled geometry, but contain holes and gaps due to the masking of stars, pixel damages or gaps between CCDs. Moreover, it has to be checked whether the assumption that all relevant angles are much smaller than the survey size is well justified for realistic configurations. In order to attain results of a more practical value, we now take no priors on the survey geometry, but retain the assumption of randomly distributed galaxy positions, and repeat the calculations of Sects. \ref{sec:estimators} and \ref{sec:covariances}.

$\Delta(\vek{\ell})$ is now used in its general form; it can be calculated at least numerically for realistic cases. The estimators (\ref{eq:estimators}) for the power spectra are kept, but will not be unbiased any more, so that we first recalculate their expectation values. Splitting the ellipticities into shear and source ellipticity terms via (\ref{eq:epsgeneral}), one arrives in case of the E-mode estimator at
\eqa{
\langle \hat{P}_{\rm E}(\bar{\ell}) \rangle &=& \frac{1}{A A_R(\bar{\ell})} \int_{A_R} \dd^2 \ell \int \dd^2 \ell_1 \int \dd^2 \ell_2 ~\Delta(\vek{\ell} - \vek{\ell_1}) \Delta^*(\vek{\ell} - \vek{\ell_2})\\ \nn
&&\times  \ba{ \tilde{\gamma}_1(\vek{\ell_1}) \tilde{\gamma}^*_1(\vek{\ell_2}) \cos^2 2\beta + \tilde{\gamma}_2(\vek{\ell_1}) \tilde{\gamma}^*_2(\vek{\ell_2}) \sin^2 2\beta + \tilde{\gamma}_1(\vek{\ell_1}) \tilde{\gamma}^*_2(\vek{\ell_2}) \sin 2\beta \cos 2\beta + \tilde{\gamma}_2(\vek{\ell_1}) \tilde{\gamma}^*_1(\vek{\ell_2}) \sin 2\beta \cos 2\beta } \\ \nn
&+& \frac{1}{\bar{n}^2 A A_R(\bar{\ell})} \int_{A_R} \dd^2 \ell \sum_{i,j} \bc{ \ba{\epsilon_{1,i}^{\rm s} \epsilon_{1,j}^{\rm s}} \cos^2 2\beta + \ba{\epsilon_{2,i}^{\rm s} \epsilon_{2,j}^{\rm s}} \sin^2 2\beta  } {\rm e}^{\ic \svek{\ell}\cdot (\svek{\theta}_i-\svek{\theta}_j)} - \frac{\sigma_\epsilon^2}{2\bar{n}}\;.
}
While the terms containing source ellipticities are processed as before and cancel with the $\sigma_\epsilon^2/2\bar{n}$ term, the shear correlators can be written in terms of convergences by means of the real and imaginary part of (\ref{eq:gamkap}) as for example
\eqa{
\ba{\tilde{\gamma}_1(\vek{\ell_1}) \tilde{\gamma}^*_1(\vek{\ell_2})} &=& \ba{ \tilde{\kappa}_{\rm E}(\vek{\ell_1}) \tilde{\kappa}^*_{\rm E}(\vek{\ell_2}) } \cos 2\beta_1 \cos 2\beta_2 + \ba{ \tilde{\kappa}_{\rm B}(\vek{\ell_1}) \tilde{\kappa}^*_{\rm B}(\vek{\ell_2}) } \sin 2\beta_1 \sin 2\beta_2\\ \nn
&-& \ba{ \tilde{\kappa}_{\rm E}(\vek{\ell_1}) \tilde{\kappa}^*_{\rm B}(\vek{\ell_2}) } \cos 2\beta_1 \sin 2\beta_2 - \ba{ \tilde{\kappa}_{\rm B}(\vek{\ell_1}) \tilde{\kappa}^*_{\rm E}(\vek{\ell_2}) } \sin 2\beta_1 \cos 2\beta_2\;, 
}
where $\beta_i$ denotes the polar angle of $\vek{\ell_i}$. One proceeds likewise for the other shear correlators. After replacing the correlators of the convergences by power spectra via (\ref{eq:Pdef}), one obtains an expectation value
\eq{
\label{eq:expPEgen}
\langle \hat{P}_{\rm E}(\bar{\ell}) \rangle = \frac{1}{A_R(\bar{\ell})} \int_{A_R(\bar{\ell})} \dd^2 \ell ~P_{\rm E}^{\rm obs}(\vek{\ell},\vek{\ell})\;,
}
where we have defined
\eq{
\label{eq:PobsE}
P_{\rm E}^{\rm obs}(\vek{\ell},\vek{\ell'})= \frac{(2\pi)^2}{A} \int \dd^2 \ell_1 ~\Delta(\vek{\ell} - \vek{\ell_1}) \Delta^*(\vek{\ell'} - \vek{\ell_1})
\bc{P_{\rm E}(\ell_1) \cos 2(\beta_1-\beta) \cos 2(\beta_1-\beta') + P_{\rm B}(\ell_1) \sin 2(\beta_1-\beta) \sin 2(\beta_1-\beta')}\;.
}
Repeating this calculation for $\langle \hat{P}_{\rm B}(\bar{\ell}) \rangle$ results in an analog to (\ref{eq:expPEgen}), now with
\eq{
P_{\rm B}^{\rm obs}(\vek{\ell},\vek{\ell'})= \frac{(2\pi)^2}{A} \int \dd^2 \ell_1 ~\Delta(\vek{\ell} - \vek{\ell_1}) \Delta^*(\vek{\ell'} - \vek{\ell_1})
\bc{P_{\rm B}(\ell_1) \cos 2(\beta_1-\beta) \cos 2(\beta_1-\beta') + P_{\rm E}(\ell_1) \sin 2(\beta_1-\beta) \sin 2(\beta_1-\beta')}\;.
}
In the special case that the assumptions of the previous sections hold, the observed power spectra are equal to the theoretical power spectrum. However, in the general case they are a convolution of $P_{\rm E}(\ell)$ and $P_{\rm B}(\ell)$ with the power spectrum of the field aperture, modified with factors stemming from the E/B-mode decomposition. This is a well-known result, derived by \citet{baumgart91} in the context of galaxy redshift surveys for arbitrary weighting functions. Similarly, we can allow $\Pi(\vek{\theta})$ not to just have values of 0 and 1, but assign to every point of the survey an individual weighting. The weighting function, i.e. in our case $\Pi(\vek{\theta})$, can then be optimized such that the variance becomes minimal. For the case of galaxy redshift surveys, \citet{feldman94} derived a minimal variance weighting under the assumption of Gaussian fluctuations; a more general consideration can be found in \citet{hamilton97}. This optimization could be used as an alternative to the numerical approach by \citet{kilbinger04}, in order to design optimal survey geometries.

The covariances of the estimators are calculated in the same manner as in Sect. \ref{sec:covariances}. After some algebra one obtains for $X=\{E,B\}$
\eq{
\ba{ \Delta P_{\rm X}(\bar{\ell}) \Delta P_{\rm X}(\bar{\ell'}) } = \frac{1}{A_R(\bar{\ell})A_R(\bar{\ell'})} \int_{A_R(\bar{\ell})} \dd^2 \ell \int_{A_R(\bar{\ell}')} \dd^2 \ell' ~2\bs{P_{\rm X}^{\rm obs}(\vek{\ell},\vek{\ell'}) + \frac{{\tilde{\sigma}}^2(\vek{\ell},\vek{\ell'})}{2\bar{n}}}^2, 
}
where ${\tilde{\sigma}}^2(\vek{\ell},\vek{\ell'}) = \bc{ \sigma_\epsilon^2 (2\pi)^2/A } ~\Delta(\vek{\ell} - \vek{\ell'}) \cos 2(\beta-\beta')$ has been defined for convenience. Now the covariance matrices are in general no longer diagonal. Furthermore, a correlation between $P_{\rm E}$ and $P_{\rm B}$ is introduced:
\eq{
\label{eq:covgen}
\langle \Delta P_{\rm E}(\bar{\ell}) \Delta P_{\rm B}(\bar{\ell'}) \rangle = \frac{1}{A_R(\bar{\ell})A_R(\bar{\ell'})} \int_{A_R(\bar{\ell})} \dd^2 \ell \int_{A_R(\bar{\ell}')} \dd^2 \ell' ~2\bs{P_{\rm cr}^{\rm obs}(\vek{\ell},\vek{\ell'}) + \frac{{\tilde{\sigma}_{\rm cr}}^2(\vek{\ell},\vek{\ell'})}{2\bar{n}}}^2,
}
where we defined
\eq{
P_{\rm cr}^{\rm obs}(\vek{\ell},\vek{\ell'}) = \frac{(2\pi)^2}{A} \int \dd^2 \ell_1 \Delta(\vek{\ell} - \vek{\ell_1}) \Delta^*(\vek{\ell'} - \vek{\ell_1})
\bc{P_{\rm E}(\ell_1) \cos 2(\beta_1-\beta) \sin 2(\beta_1-\beta') - P_{\rm B}(\ell_1) \sin 2(\beta_1-\beta) \cos 2(\beta_1-\beta')}
}
and ${\tilde{\sigma}_{\rm cr}}^2(\vek{\ell},\vek{\ell'}) = \bc{ \sigma_\epsilon^2 (2\pi)^2/A} ~\Delta(\vek{\ell} - \vek{\ell'}) \sin 2(\beta-\beta')$. Note that in the limit $\Delta(\vek{\ell}) \rightarrow \delta^{(2)}(\vek{\ell})$, the foregoing results for a simple survey geometry are regained also for the covariances.

\subsection{Some specific examples}

\begin{figure}[t]
\begin{minipage}[c]{0.58\textwidth}
\centering
\includegraphics[scale=.72]{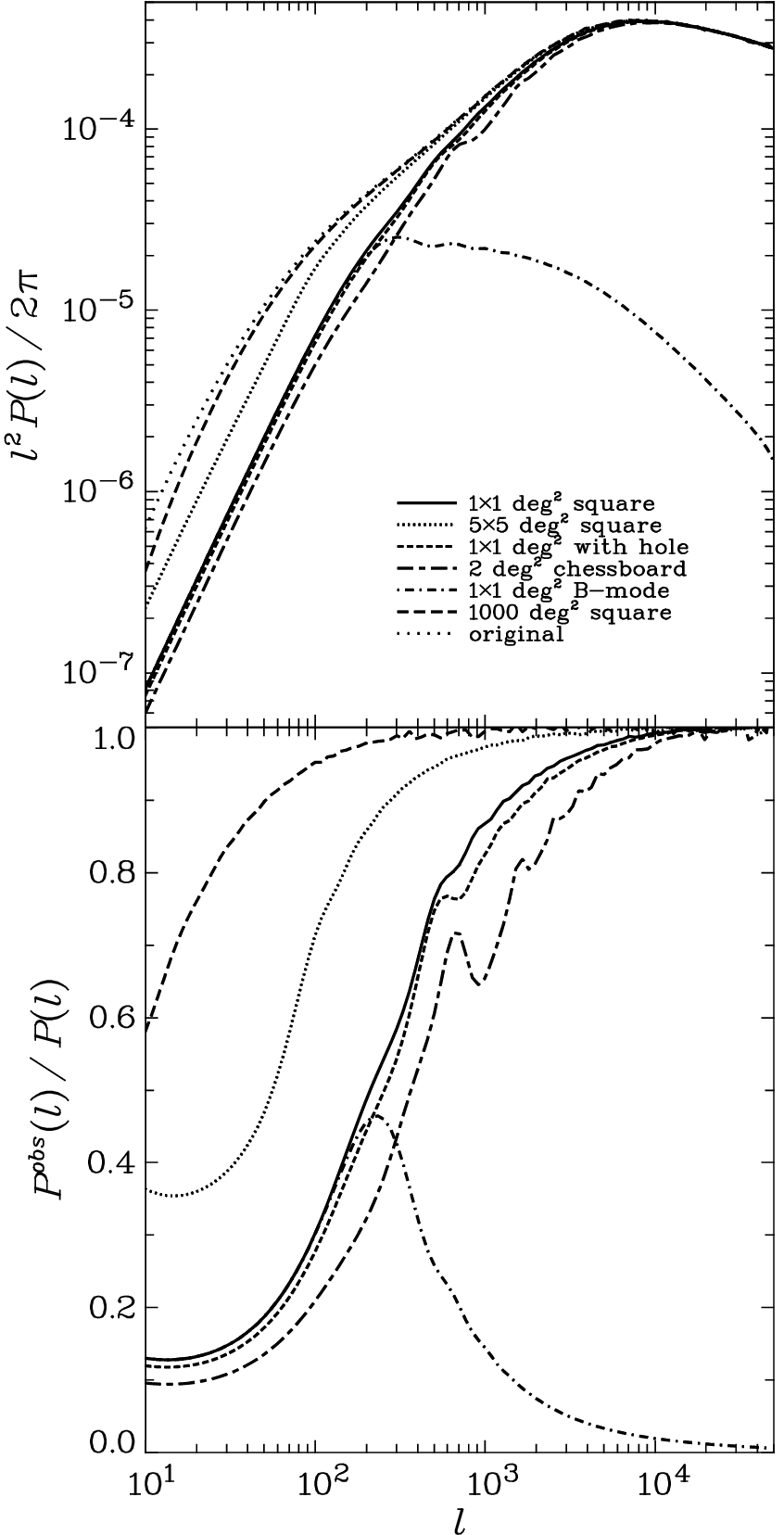}
\end{minipage}% 
\begin{minipage}[c]{0.42\textwidth}
\centering
\includegraphics[scale=.57]{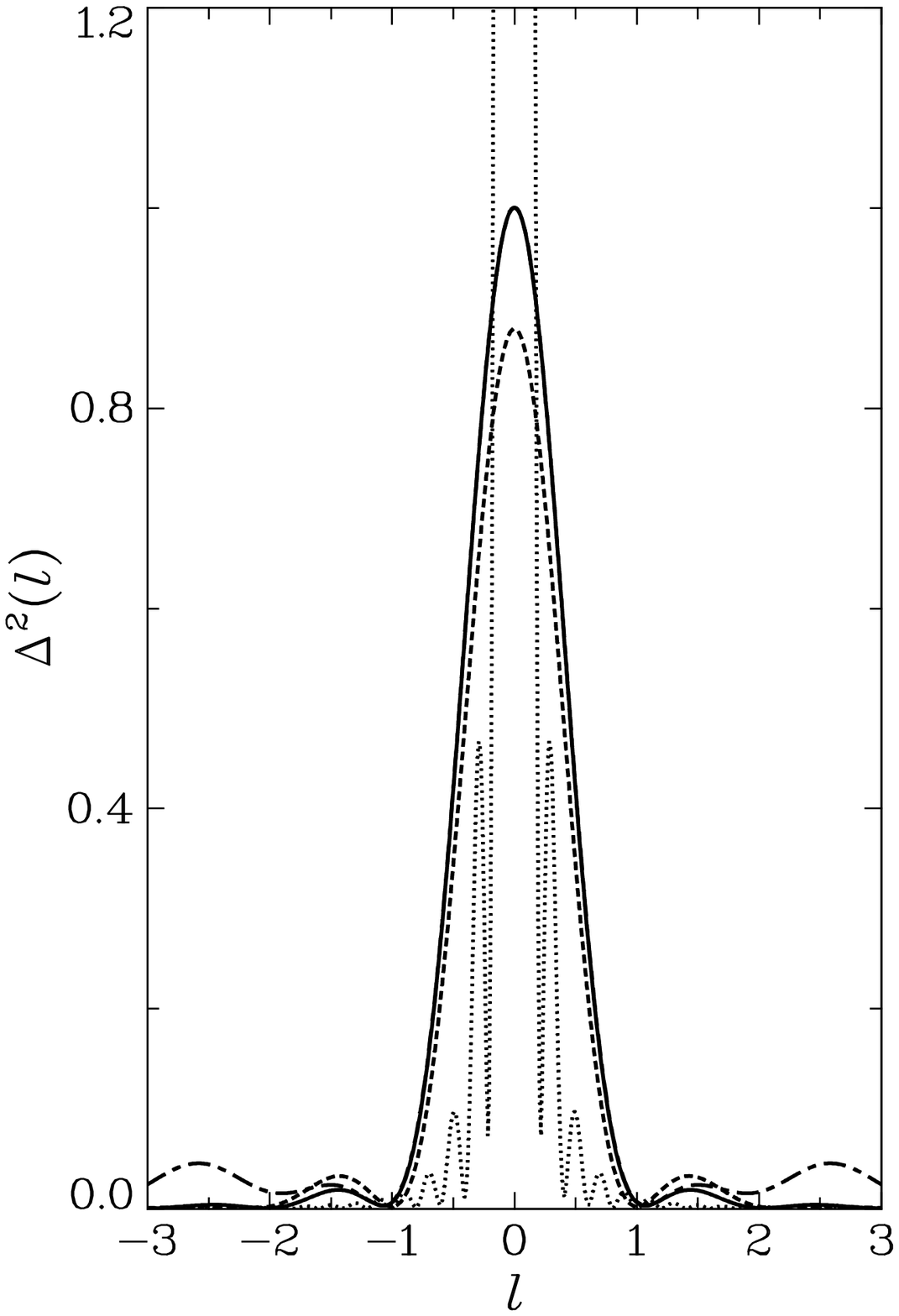}
\caption{Influence of survey geometry on the estimated power spectrum. \textit{Upper left panel}: E-mode power spectra $\langle P_{\rm E}^{\rm obs} \rangle (\ell)$, averaged over the polar angle of $\vek{\ell}$, obtained for different survey geometries. The sparsely dotted curve shows the original E-mode power spectrum $P_{\rm E}(\ell)$. The solid line corresponds to a square field with 1$\times$1$\,$deg$^2$, the dotted one to a square field with 5$\times$5$\,$deg$^2$ and the long-dashed one to a square survey with 1000$\,{\rm deg}^2$. In the case of the short-dashed curve the 1$\,$deg$^2$-field contains a hole, while the dot-long-dashed line corresponds to the chessboard configuration. For comparison, the corresponding observed B-mode power spectrum (caused by E-modes) for the 1$\times$1$\,$deg$^2$ square field geometry is plotted as dot-short-dashed curve. \textit{Lower left panel}: Fraction of observed E-mode power spectrum over original E-mode power spectrum. The coding is identical to the left upper panel. \textit{Right panel}: Power spectrum of the phase-averaged field aperture $\langle \Delta^2 \rangle (\ell)$. The coding of the curves is the same as in the left panels, the curve for the chessboard configuration lying on top of the solid line in the central part of the diagram. The axes are normalized such that $\langle \Delta^2 \rangle (0)=1$ for the 1$\,$deg$^2$-field. Note that the maximum of the dotted curve is at 25. The even narrower curve for the 1000$\,{\rm deg}^2$ field has not been plotted for reasons of clarity.
} 
\label{fig:geometry}
\end{minipage}
\end{figure} 

In order to demonstrate the influence of field geometry on our estimators, we consider a number of examples for which $\Delta(\vek{\ell})$ can be calculated analytically. First, we assume simple square fields of size $1\times1 ~\rm{deg}^2$ and $5\times5 ~\rm{deg}^2$ each, as well as a $1000\,{\rm deg}^2$ square geometry, its size being of the order of currently planned cosmic shear surveys. Second, we place a hole of $0.25\times 0.25 ~\rm{deg}^2$ at the center of the 1$\,$deg$^2$ survey. Finally, a chessboard configuration is examined, where 8 sub-fields of size $0.125 ~\rm{deg}^2$ each are embedded in a 2$\,$deg$^2$ field to get the same effective area as for the 1$\,$deg$^2$ survey in the first example. By letting $\Delta \ell \rightarrow 0$ in (\ref{eq:expPEgen}), we calculate a phase-averaged observed power spectrum, $\langle P_{\rm X}^{\rm obs} \rangle (\ell) = 1/2\pi \int_0^{2\pi} \!\dd \beta \;P_{\rm X}^{\rm obs}(\vek{\ell},\vek{\ell})$, for each geometry. We set $P_{\rm B}(\ell) \equiv 0$; however, $P_{\rm B}^{\rm obs}$ does not vanish since it has contributions from $P_{\rm E}(\ell)$, see (\ref{eq:PobsE}).

In Fig. \ref{fig:geometry}, left upper panel, the resulting phase-averaged observed E-mode power spectra for all field configurations are plotted. In addition, the theoretical power spectrum as well as the observed B-mode power spectrum in the case of the 1$\,$deg$^2$ square field are depicted. The lower left panel shows the amplitude of the observed E-mode power spectra relative to the theoretical one. Moreover, the aperture power spectra $\langle \Delta^2 \rangle (\ell)$, again averaged over $\beta$, for the different geometries are given in the right panel, for our examples all being composed of sinc$^2$-functions. Note that our assumptions about the survey geometry made in the previous sections would lead to a Dirac-delta spike at $\ell=0$ in this plot.

For large $\ell$ all observed power spectra trace the theoretical one well, whereas the configurations with small survey area deviate significantly for $\ell \apprle 1000$. Due to the limited field extent no ellipticity pairs are measured at large separations on the sky, corresponding to a lack of information about the power at small $\ell$. In the case of larger effective survey size, deviations are therefore smaller and start at smaller angular frequencies. The relative amplitude of the observed power spectrum decreases below $90\,\%$ at $\ell \apprle 300$ for a survey size of $5\times5\,\rm{deg}^2$, and at $\ell \apprle 50$ for the $1000\,\rm{deg}^2$ survey. The convolution kernel of the $5\times5\,$deg$^2$-field is already quite narrow, the Dirac-delta distribution therefore being a reasonable approximation in the case of large survey areas.

Cutting a hole into the field has very little effect on the observed power as there is still information available on all scales; only the noise is expected to increase. The relative height of the side lobes of the aperture power spectrum increases moderately, so that $\langle P_{\rm E}^{\rm obs} \rangle (\ell)$ is slightly lowered. On the contrary, the curve for the chessboard configuration lies significantly lower than for the 1$\,$deg$^2$ field. This can be explained by the pronounced side lobes of the otherwise identical aperture power spectrum, which are generated by the periodicity of the field geometry. Power is already reduced considerably on angular scales corresponding to the size of the chessboard squares, leading to an early deviation from the theoretical power spectrum.

As expected, the observed B-mode power spectrum is significantly smaller than the E-mode signals for larger angular frequencies, but for $\ell \apprle 200$, corresponding to angular scales which do not fit into the field, the B-modes attain the same order of magnitude as the observed E-modes because in this range the phase-averaging of the $\sin^2 2(\beta_1-\beta)$ and the $\cos^2 2(\beta_1-\beta)$ term in $P_{\rm E}^{\rm obs}$ yields the same result or, illustratively, E- and B-modes cannot be distinguished anymore on large angular scales. This means that apart from a bias one also has to take into account a leakage between E- and B-modes, which is severe in case one would like to use the power spectrum estimator (\ref{eq:estimators}) to measure cosmological B-modes. Since these are potentially significantly smaller than the E-mode signal, even a small leakage will then cause a strong contamination.

Taking into account the results for these exemplary field geometries, it is unfortunately not so obvious how the calculation of the covariances of other second-order measures should proceed in the general case, using (\ref{eq:covgen}), since (\ref{eq:filters}) has to hold also for the estimators, which is not the case if $\hat{P}_{\rm E}(\bar{\ell})$ is biased. However, the survey geometries considered here are quite artificial; in the real world, one will probably be faced with a large contiguous field, with certain regions, located quasi randomly, been cut out. In this situation, the covariance is expected to behave like in the connected field geometry case, with the survey area $A$ taken to be the effective area, i.e. with the cut-out regions subtracted -- our case with the hole supports this conjecture. The results for the specific examples discussed above may then be used to assess the accuracy of this approximation.

\section{Conclusions}
\label{sec:conclusions}

In this paper we presented a new approach to calculate analytically covariance matrices of second-order cosmic shear measures under the assumption of Gaussian density fluctuations, uniformly distributed galaxies and a simple survey geometry. By introducing estimators for the E- and B-mode power spectra and performing the derivation in Fourier space, we arrived at compact results which will simplify further analytical considerations and speed up numerical work. These results are fully equivalent to those found in a real-space approach by Paper I. 

We then formulated the covariance matrices in the context of shear tomography. Compared to Monte Carlo simulations employed by \citet{simon04}, the analytical ansatz leads to an enormous gain in computational speed. However, the former method has the advantage that it is directly applicable to arbitrary survey geometries.

We also generalized our findings to arbitrary survey geometries and examined the bias thereby introduced on our estimators. In this context one might pose the question why the convergence power spectrum, which contains all second-order information about cosmic shear, is not directly estimated from the data and solely used for further analysis. However, the power spectrum estimators incorporate a Fourier transformation for which galaxy ellipticities have to be measured in principle on the whole sky. This problem is avoided by the assumption $|\vek{\theta}|^2 \ll A$, but will lead to a bias in case of finite and more complex field geometries. In principle, the power spectrum could also be calculated from real-space measures via the inverse of (\ref{eq:filters}), but this is hindered by the infinite range of integration over the angular separation $\theta$. 

Although the theoretical power spectrum cannot be inferred to
satisfactory accuracy from the estimators (\ref{eq:estimators}), they
may still be of use in cosmic shear data analysis. As
demonstrated in Sect. \ref{sec:general}, it is relatively
straightforward to calculate expectation values of observed power
spectra $P_{\rm X}^{\rm obs}$ from the theoretical $P_{\rm X}(\ell)$ and
successively compare these to results from observations. Another way to 
directly extract the power spectrum was discussed in \citet{hu01} who determine
band powers with a maximum likelihood method, which is able to recover
the power spectrum irrespective of the survey geometry, but is not
suited for large fields due to its computational
requirements. Furthermore, due to its narrow filter, $M^2_{\rm ap}$ can
yield a fairly accurate estimate of the power spectrum localized at scales 
$\ell\approx 4.25/\theta$ \citep{bartelmann99}. In order to determine cosmological parameters though, one would not make use of the power spectrum. For this purpose the correlation functions are the second-order measures of choice since they remain unbiased for arbitrary survey geometries.

All methods discussed here are limited by the assumption of
Gaussianity, that will break down on small, non-linear
scales. Anyway, in case of non-Gaussian density fluctuations the power
spectrum is not sufficient for their characterization any more, so
higher-order statistics have to be taken into account. In fact, as
shown by \citet{kilbinger05} and \citet{semboloni07},
the non-Gaussianity modifies the covariance substantially on angular
scales below $\sim 15'$, depending on the source redshift
distribution. The increase of the covariance due to non-Gaussianity is
so strong that the shape noise term is nearly always sub-dominant
compared to the cosmic variance contribution. \citet{semboloni07} have used 
numerical simulations to quantify the deviation of
the covariance matrix of the shear correlation function from the
Gaussian case, giving analytic fit functions. We hypothesize that a
similar approach is possible in the context of our formulation, where
now the fitted correction function for non-Gaussianity will be be
applied to the power spectrum covariance (\ref{eq:covP1}). We will
consider this in future work.

For the full treatment of four-point correlators needed for the covariances 
of second-order statistics in the non-Gaussian regime, only few analytical 
methods exist, so that one has to rely on numerical work, for example
ray-tracing simulations. As also the analytical fit functions for the
input power spectra \citep{PeacockDodds,smith03} have
limited accuracy, simulations will be indispensable for the
determination of covariances for future cosmic shear surveys, designed
to serve as high-precision tools for cosmology.

Moreover, the calculation of covariances via a power spectrum approach has revealed an interesting analogy to the analysis of the Cosmic Microwave Background. Similarly, power spectrum estimators for the E- and B-mode polarization of the CMB are defined in an idealized survey geometry, namely on the full sky. Due to incomplete sky coverage or masking of the Milky Way in real data these estimators become biased and mix E- and B-modes. However, in the context of CMB polarization there exist methods which allow to debias estimators and single out B-modes, see e.g. \citet{hivon02} and \citet{smith06b,smith06}. The further investigation of these methods, transfered to cosmic shear, may prove valuable.

\begin{acknowledgements}
We thank Patrick Simon, Martin Kilbinger and Jan Hartlap for useful
discussion. Furthermore, we are grateful to Antony Lewis who pointed us to the analogy to CMB analysis, and to our referee for his very helpful comments. This work was supported by the Deutsche
Forschungsgemeinschaft under the project SCHN 342/6--1, the
DFG-Priority Programme 1177, and the Transregio-33 `The Dark
Universe'.
\end{acknowledgements}

\bibliographystyle{aa}
%\bibliography{../../diplom/tex/bibliography}

\end{document}